\documentclass[longauth]{aa}  

\usepackage{booktabs}
\usepackage{graphicx}
\usepackage{txfonts}
\usepackage[]{hyperref}
\usepackage{orcidlink}
\usepackage[switch]{lineno} 

\newcounter{pta}
\renewcommand{\thepta}{\Roman{pta}} 

\DeclareRobustCommand{\defpta}[1]{%
   \refstepcounter{pta}%
   \thepta\label{#1}%
}

\newcommand{\refpta}[1]{\ref{#1}}

\begin{document} 

   \title{The second data release from the European Pulsar Timing Array}
   \subtitle{II. Customised pulsar noise models for spatially correlated gravitational waves}
\def\wm{2} 
   
\author{
    J.~Antoniadis\orcidlink{0000-0003-4453-776}\inst{\ref{forth},\ref{mpifr}\ifnum\wm>1,\refpta{epta}\fi},
    \ifnum\wm>1 P.~Arumugam\orcidlink{0000-0001-9264-8024}\inst{\ref{IITR}\ifnum\wm>1,\refpta{inpta}\fi},\fi
    \ifnum\wm>1 S.~Arumugam\orcidlink{0009-0001-3587-6622}\inst{\ref{IITH_El}\ifnum\wm>1,\refpta{inpta}\fi},\fi
    S.~Babak\orcidlink{0000-0001-7469-4250}\inst{\ref{apc}\ifnum\wm>1,\refpta{epta}\fi},
    \ifnum\wm>1 M.~Bagchi\orcidlink{0000-0001-8640-8186}\inst{\ref{IMSc},\ref{HBNI}\ifnum\wm>1,\refpta{inpta}\fi},\fi
    A.-S.~Bak~Nielsen\orcidlink{ 0000-0002-1298-9392}\inst{\ref{mpifr},\ref{unibi}\ifnum\wm>1,\refpta{epta}\fi},
    C.~G.~Bassa\orcidlink{0000-0002-1429-9010}\inst{\ref{astron}\ifnum\wm>1,\refpta{epta}\fi},
    \ifnum\wm>1 A.~Bathula\orcidlink{0000-0001-7947-6703} \inst{\ref{IISERM}\ifnum\wm>1,\refpta{inpta}\fi},\fi
    A.~Berthereau\inst{\ref{lpc2e},\ref{nancay}\ifnum\wm>1,\refpta{epta}\fi},
    M.~Bonetti\orcidlink{0000-0001-7889-6810}\inst{\ref{unimib},\ref{infn-unimib},\ref{inaf-brera}\ifnum\wm>1,\refpta{epta}\fi},
    E.~Bortolas\inst{\ref{unimib},\ref{infn-unimib},\ref{inaf-brera}\ifnum\wm>1,\refpta{epta}\fi},
    P.~R.~Brook\orcidlink{0000-0003-3053-6538}\inst{\ref{unibir}\ifnum\wm>1,\refpta{epta}\fi},
    M.~Burgay\orcidlink{0000-0002-8265-4344}\inst{\ref{inaf-oac}\ifnum\wm>1,\refpta{epta}\fi},
    R.~N.~Caballero\orcidlink{0000-0001-9084-9427}\inst{\ref{HOU}\ifnum\wm>1,\refpta{epta}\fi},
    A.~Chalumeau\orcidlink{0000-0003-2111-1001}\inst{\ref{unimib}\ifnum\wm>1,\refpta{epta}\fi}\ifnum\wm=2\thanks{aurelien.chalumeau@unimib.it}\fi,
    D.~J.~Champion\orcidlink{0000-0003-1361-7723}\inst{\ref{mpifr}\ifnum\wm>1,\refpta{epta}\fi},
    S.~Chanlaridis\orcidlink{0000-0002-9323-9728}\inst{\ref{forth}\ifnum\wm>1,\refpta{epta}\fi},
    S.~Chen\orcidlink{0000-0002-3118-5963}\inst{\ref{kiaa}\ifnum\wm>1,\refpta{epta}\fi}\ifnum\wm=3\thanks{sychen@pku.edu.cn}\fi,
    I.~Cognard\orcidlink{0000-0002-1775-9692}\inst{\ref{lpc2e},\ref{nancay}\ifnum\wm>1,\refpta{epta}\fi},
    \ifnum\wm>1 S.~Dandapat\orcidlink{0000-0003-4965-9220}\inst{\ref{TIFR}\ifnum\wm>1,\refpta{inpta}\fi},\fi
    \ifnum\wm>1 D.~Deb\orcidlink{0000-0003-4067-5283}\inst{\ref{IMSc}\ifnum\wm>1,\refpta{inpta}\fi},      \fi
    \ifnum\wm>1 S.~Desai\orcidlink{0000-0002-0466-3288}\inst{\ref{IITH_Ph}\ifnum\wm>1,\refpta{inpta}\fi},\fi
    G.~Desvignes\orcidlink{0000-0003-3922-4055}\inst{\ref{mpifr}\ifnum\wm>1,\refpta{epta}\fi},
    \ifnum\wm>1 N.~Dhanda-Batra \inst{\ref{UoD}\ifnum\wm>1,\refpta{inpta}\fi},\fi
    \ifnum\wm>1 C.~Dwivedi\orcidlink{0000-0002-8804-650X}\inst{\ref{IIST}\ifnum\wm>1,\refpta{inpta}\fi},\fi
    M.~Falxa\inst{\ref{apc},\ref{lpc2e}\ifnum\wm>1,\refpta{epta}\fi}
    R.~D.~Ferdman\inst{\ref{uea}\ifnum\wm>1,\refpta{epta}\fi},
    A.~Franchini\orcidlink{0000-0002-8400-0969}\inst{\ref{unimib},\ref{infn-unimib}\ifnum\wm>1,\refpta{epta}\fi},
    J.~R.~Gair\orcidlink{0000-0002-1671-3668}\inst{\ref{aei}\ifnum\wm>1,\refpta{epta}\fi},
    B.~Goncharov\orcidlink{0000-0003-3189-5807}\inst{\ref{gssi},\ref{lngs}\ifnum\wm>1,\refpta{epta}\fi}
    \ifnum\wm>1 A.~Gopakumar\orcidlink{0000-0003-4274-4369}\inst{\ref{TIFR}\ifnum\wm>1,\refpta{inpta}\fi},\fi
    E.~Graikou\inst{\ref{mpifr}\ifnum\wm>1,\refpta{epta}\fi},
    J.-M.~Grie{\ss}meier\orcidlink{0000-0003-3362-7996}\inst{\ref{lpc2e},\ref{nancay}\ifnum\wm>1,\refpta{epta}\fi},
    L.~Guillemot\orcidlink{0000-0002-9049-8716}\inst{\ref{lpc2e},\ref{nancay}\ifnum\wm>1,\refpta{epta}\fi},
    Y.~J.~Guo\inst{\ref{mpifr}\ifnum\wm>1,\refpta{epta}\fi}\ifnum\wm=3\thanks{yjguo@mpifr-bonn.mpg.de}\fi,
    \ifnum\wm>1 Y.~Gupta\orcidlink{0000-0001-5765-0619}\inst{\ref{NCRA}\ifnum\wm>1,\refpta{inpta}\fi},\fi
    \ifnum\wm>1 S.~Hisano\orcidlink{0000-0002-7700-3379}\inst{\ref{KU_J}\ifnum\wm>1,\refpta{inpta}\fi},\fi
    H.~Hu\orcidlink{0000-0002-3407-8071}\inst{\ref{mpifr}\ifnum\wm>1,\refpta{epta}\fi}, 
    F.~Iraci\inst{\ref{unica}\ref{inaf-oac}\ifnum\wm>1,\refpta{epta}\fi},
    D.~Izquierdo-Villalba\orcidlink{0000-0002-6143-1491}\inst{\ref{unimib},\ref{infn-unimib}\ifnum\wm>1,\refpta{epta}\fi},
    J.~Jang\orcidlink{0000-0003-4454-0204}\inst{\ref{mpifr}\ifnum\wm>1,\refpta{epta}\fi}\ifnum\wm=1\thanks{jjang@mpifr-bonn.mpg.de}\fi,
    J.~Jawor\orcidlink{0000-0003-3391-0011}\inst{\ref{mpifr}\ifnum\wm>1,\refpta{epta}\fi},
    G.~H.~Janssen\orcidlink{0000-0003-3068-3677}\inst{\ref{astron},\ref{imapp}\ifnum\wm>1,\refpta{epta}\fi},
    A.~Jessner\orcidlink{0000-0001-6152-9504}\inst{\ref{mpifr}\ifnum\wm>1,\refpta{epta}\fi},
    \ifnum\wm>1 B.~C.~Joshi\orcidlink{0000-0002-0863-7781}\inst{\ref{NCRA},\ref{IITR}\ifnum\wm>1,\refpta{inpta}\fi},\fi
    \ifnum\wm>1 F.~Kareem\orcidlink{0000-0003-2444-838X} \inst{\ref{IISERK},\ref{CESSI}\ifnum\wm>1,\refpta{inpta}\fi},\fi
    R.~Karuppusamy\orcidlink{0000-0002-5307-2919}\inst{\ref{mpifr}\ifnum\wm>1,\refpta{epta}\fi},
    E.~F.~Keane\orcidlink{0000-0002-4553-655X}\inst{\ref{tcd}\ifnum\wm>1,\refpta{epta}\fi},
    M.~J.~Keith\orcidlink{0000-0001-5567-5492}\inst{\ref{jbca}\ifnum\wm>1,\refpta{epta}\fi}\ifnum\wm=2\thanks{michael.keith@manchester.ac.uk}\fi,
    \ifnum\wm>1 D.~Kharbanda\orcidlink{0000-0001-8863-4152}\inst{\ref{IITH_Ph}\ifnum\wm>1,\refpta{inpta}\fi},\fi
    \ifnum\wm>1 T.~Kikunaga\orcidlink{0000-0002-5016-3567} \inst{\ref{KU_J}\ifnum\wm>1,\refpta{inpta}\fi},\fi
    \ifnum\wm>1 N.~Kolhe\orcidlink{0000-0003-3528-9863} \inst{\ref{XCM}\ifnum\wm>1,\refpta{inpta}\fi},\fi
    M.~Kramer\inst{\ref{mpifr},\ref{jbca}\ifnum\wm>1,\refpta{epta}\fi},
    M.~A.~Krishnakumar\orcidlink{0000-0003-4528-2745}\inst{\ref{mpifr},\ref{unibi}\ifnum\wm>1,\refpta{epta}\fi\ifnum\wm>1,\refpta{inpta}\fi},
    K.~Lackeos\orcidlink{0000-0002-6554-3722}\inst{\ref{mpifr}\ifnum\wm>1,\refpta{epta}\fi},
    K.~J.~Lee\inst{3,8,\ref{mpifr}\ifnum\wm>1,\refpta{epta}\fi},
    K.~Liu\inst{\ref{mpifr}\ifnum\wm>1,\refpta{epta}\fi}\ifnum\wm=1\thanks{kliu@mpifr-bonn.mpg.de}\fi,
    Y.~Liu\orcidlink{0000-0001-9986-9360}\inst{\ref{naoc}, \ref{unibi}\ifnum\wm>1,\refpta{epta}\fi},
    A.~G.~Lyne\inst{\ref{jbca}\ifnum\wm>1,\refpta{epta}\fi},
    J.~W.~McKee\orcidlink{0000-0002-2885-8485}\inst{\ref{milne},\ref{daim}\ifnum\wm>1,\refpta{epta}\fi},
    \ifnum\wm>1 Y.~Maan\inst{\ref{NCRA}\ifnum\wm>1,\refpta{inpta}\fi},\fi
    R.~A.~Main\inst{\ref{mpifr}\ifnum\wm>1,\refpta{epta}\fi},
    M.~B.~Mickaliger\orcidlink{0000-0001-6798-5682}\inst{\ref{jbca}\ifnum\wm>1,\refpta{epta}\fi},
    I.~C.~Ni\c{t}u\orcidlink{0000-0003-3611-3464}\inst{\ref{jbca}\ifnum\wm>1,\refpta{epta}\fi},
    \ifnum\wm>1 K.~Nobleson\orcidlink{0000-0003-2715-4504}\inst{\ref{BITS}\ifnum\wm>1,\refpta{inpta}\fi},\fi
    \ifnum\wm>1 A.~K.~Paladi\orcidlink{0000-0002-8651-9510}\inst{\ref{IISc}\ifnum\wm>1,\refpta{inpta}\fi},\fi
    A.~Parthasarathy\orcidlink{0000-0002-4140-5616}\inst{\ref{mpifr}\ifnum\wm>1,\refpta{epta}\fi}\ifnum\wm=2\thanks{aparthas@mpifr-bonn.mpg.de}\fi,
    B.~B.~P.~Perera\orcidlink{0000-0002-8509-5947}\inst{\ref{arecibo}\ifnum\wm>1,\refpta{epta}\fi},
    D.~Perrodin\orcidlink{0000-0002-1806-2483}\inst{\ref{inaf-oac}\ifnum\wm>1,\refpta{epta}\fi},
    A.~Petiteau\orcidlink{0000-0002-7371-9695}\inst{\ref{irfu},\ref{apc}\ifnum\wm>1,\refpta{epta}\fi},
    N.~K.~Porayko\inst{\ref{unimib},\ref{mpifr}\ifnum\wm>1,\refpta{epta}\fi},
    A.~Possenti\inst{\ref{inaf-oac}\ifnum\wm>1,\refpta{epta}\fi},
    \ifnum\wm>1 T.~Prabu\inst{\ref{RRI}\ifnum\wm>1,\refpta{inpta}\fi},\fi
    H.~Quelquejay~Leclere\inst{\ref{apc}\ifnum\wm>1,\refpta{epta}\fi}
    \ifnum\wm>1 P.~Rana\orcidlink{0000-0001-6184-5195}\inst{\ref{TIFR}\ifnum\wm>1,\refpta{inpta}\fi},\fi
    A.~Samajdar\orcidlink{0000-0002-0857-6018}\inst{\ref{uni-potsdam}\ifnum\wm>1,\refpta{epta}\fi},
    S.~A.~Sanidas\inst{\ref{jbca}\ifnum\wm>1,\refpta{epta}\fi},
    A.~Sesana\inst{\ref{unimib},\ref{infn-unimib},\ref{inaf-brera}\ifnum\wm>1,\refpta{epta}\fi},
    G.~Shaifullah\orcidlink{0000-0002-8452-4834}\inst{\ref{unimib},\ref{infn-unimib},\ref{inaf-oac}\ifnum\wm>1,\refpta{epta}\fi}\ifnum\wm=1\thanks{golam.shaifullah@unimib.it}\fi,
    \ifnum\wm>1 J.~Singha\orcidlink{0000-0002-1636-9414}\inst{\ref{IITR}\ifnum\wm>1,\refpta{inpta}\fi},\fi
    L.~Speri\orcidlink{0000-0002-5442-7267}\inst{\ref{aei}\ifnum\wm>1,\refpta{epta}\fi},
    R.~Spiewak\inst{\ref{jbca}\ifnum\wm>1,\refpta{epta}\fi},
    \ifnum\wm>1 A.~Srivastava\orcidlink{0000-0003-3531-7887} \inst{\ref{IITH_Ph}\ifnum\wm>1,\refpta{inpta}\fi},\fi
    B.~W.~Stappers\inst{\ref{jbca}\ifnum\wm>1,\refpta{epta}\fi},
    \ifnum\wm>1 M.~Surnis\orcidlink{0000-0002-9507-6985}\inst{\ref{IISERB}\ifnum\wm>1,\refpta{inpta}\fi},\fi
    S.~C.~Susarla\orcidlink{0000-0003-4332-8201}\inst{\ref{uog}\ifnum\wm>1,\refpta{epta}\fi},
    \ifnum\wm>1 A.~Susobhanan\orcidlink{0000-0002-2820-0931}\inst{\ref{CGCA}\ifnum\wm>1,\refpta{inpta}\fi},\fi
    \ifnum\wm>1 K.~Takahashi\orcidlink{0000-0002-3034-5769}\inst{\ref{KU_J1},\ref{KU_J2}\ifnum\wm>1,\refpta{inpta}\fi} \fi
    \ifnum\wm>1 P.~Tarafdar\orcidlink{0000-0001-6921-4195}\inst{\ref{IMSc}\ifnum\wm>1,\refpta{inpta}\fi}\fi
    G.~Theureau\orcidlink{0000-0002-3649-276X}\inst{\ref{lpc2e}, \ref{nancay}, \ref{luth}\ifnum\wm>1,\refpta{epta}\fi},
    C.~Tiburzi\inst{\ref{inaf-oac}\ifnum\wm>1,\refpta{epta}\fi},
    E.~van~der~Wateren\orcidlink{0000-0003-0382-8463}\inst{\ref{astron},\ref{imapp}\ifnum\wm>1,\refpta{epta}\fi},
    A.~Vecchio\orcidlink{0000-0002-6254-1617}\inst{\ref{unibir}\ifnum\wm>1,\refpta{epta}\fi},
    V.~Venkatraman~Krishnan\orcidlink{0000-0001-9518-9819}\inst{\ref{mpifr}\ifnum\wm>1,\refpta{epta}\fi},
    J.~P.~W.~Verbiest\orcidlink{0000-0002-4088-896X}\inst{\ref{FSI},\ref{unibi},\ref{mpifr}\ifnum\wm>1,\refpta{epta}\fi},
    J.~Wang\orcidlink{0000-0003-1933-6498}\inst{\ref{unibi}, \ref{airub}, \ref{bnuz}\ifnum\wm>1,\refpta{epta}\fi},
    L.~Wang\inst{\ref{jbca}\ifnum\wm>1,\refpta{epta}\fi} and 
    Z.~Wu\orcidlink{0000-0002-1381-7859}\inst{\ref{naoc},\ref{unibi}\ifnum\wm>1,\refpta{epta}\fi}.
    }

\institute{
{Institute of Astrophysics, FORTH, N. Plastira 100, 70013, Heraklion, Greece\label{forth}}\and 
{Max-Planck-Institut f{\"u}r Radioastronomie, Auf dem H{\"u}gel 69, 53121 Bonn, Germany\label{mpifr}}\and
\ifnum\wm>1{Department of Physics, Indian Institute of Technology Roorkee, Roorkee-247667, India\label{IITR}}\and \fi
\ifnum\wm>1{Department of Electrical Engineering, IIT Hyderabad, Kandi, Telangana 502284, India \label{IITH_El}}\and\fi
{Universit{\'e} Paris Cit{\'e}, CNRS, Astroparticule et Cosmologie, 75013 Paris, France\label{apc}}\and
\ifnum\wm>1{The Institute of Mathematical Sciences, C. I. T. Campus, Taramani, Chennai 600113, India \label{IMSc}}\and\fi
\ifnum\wm>1{Homi Bhabha National Institute, Training School Complex, Anushakti Nagar, Mumbai 400094, India \label{HBNI}}\and\fi
{Fakult{\"a}t f{\"u}r Physik, Universit{\"a}t Bielefeld, Postfach 100131, 33501 Bielefeld, Germany\label{unibi}}\and
{ASTRON, Netherlands Institute for Radio Astronomy, Oude Hoogeveensedijk 4, 7991 PD, Dwingeloo, The Netherlands\label{astron}}\and
\ifnum\wm>1{Department of Physical Sciences, Indian Institute of Science Education and Research, Mohali, Punjab 140306, India \label{IISERM}}\and\fi
{Laboratoire de Physique et Chimie de l'Environnement et de l'Espace, Universit\'e d'Orl\'eans / CNRS, 45071 Orl\'eans Cedex 02, France \label{lpc2e}}\and
{Observatoire Radioastronomique de Nan\c{c}ay, Observatoire de Paris, Universit\'e PSL, Université d'Orl\'eans, CNRS, 18330 Nan\c{c}ay, France\label{nancay}}\and
{Dipartimento di Fisica ``G. Occhialini", Universit{\'a} degli Studi di Milano-Bicocca, Piazza della Scienza 3, I-20126 Milano, Italy\label{unimib}}\and
{INFN, Sezione di Milano-Bicocca, Piazza della Scienza 3, I-20126 Milano, Italy\label{infn-unimib}}\and
{INAF - Osservatorio Astronomico di Brera, via Brera 20, I-20121 Milano, Italy\label{inaf-brera}}\and
{Institute for Gravitational Wave Astronomy and School of Physics and Astronomy, University of Birmingham, Edgbaston, Birmingham B15 2TT, UK\label{unibir}}\and
{INAF - Osservatorio Astronomico di Cagliari, via della Scienza 5, 09047 Selargius (CA), Italy\label{inaf-oac}}\and
{Hellenic Open University, School of Science and Technology, 26335 Patras, Greece\label{HOU}}\and
{Kavli Institute for Astronomy and Astrophysics, Peking University, Beijing 100871, P. R. China\label{kiaa}}\and
\ifnum\wm>1{Department of Astronomy and Astrophysics, Tata Institute of Fundamental Research, Homi Bhabha Road, Navy Nagar, Colaba, Mumbai 400005, India \label{TIFR}}\and\fi
\ifnum\wm>1{Department of Physics, IIT Hyderabad, Kandi, Telangana 502284, India \label{IITH_Ph}}\and\fi
\ifnum\wm>1{Department of Physics and Astrophysics, University of Delhi, Delhi 110007, India \label{UoD}}\and\fi
\ifnum\wm>1{Department of Earth and Space Sciences, Indian Institute of Space Science and Technology, Valiamala, Thiruvananthapuram, Kerala 695547,India \label{IIST}}\and\fi
{School of Physics, Faculty of Science, University of East Anglia, Norwich NR4 7TJ, UK\label{uea}}\and
{Max Planck Institute for Gravitational Physics (Albert Einstein Institute), Am Mu{\"u}hlenberg 1, 14476 Potsdam, Germany\label{aei}}\and
{Gran Sasso Science Institute (GSSI), I-67100 L'Aquila, Italy \label{gssi}}\and
{INFN, Laboratori Nazionali del Gran Sasso, I-67100 Assergi, Italy \label{lngs}}\and 
\ifnum\wm>1{National Centre for Radio Astrophysics, Pune University Campus, Pune 411007, India \label{NCRA}}\and\fi
\ifnum\wm>1{Kumamoto University, Graduate School of Science and Technology, Kumamoto, 860-8555, Japan \label{KU_J}}\and\fi
{Universit{\'a} di Cagliari, Dipartimento di Fisica, S.P. Monserrato-Sestu Km 0,700 - 09042 Monserrato (CA), Italy\label{unica}}\and
{Department of Astrophysics/IMAPP, Radboud University Nijmegen, P.O. Box 9010, 6500 GL Nijmegen, The Netherlands\label{imapp}}\and
\ifnum\wm>1{Department of Physical Sciences,Indian Institute of Science Education and Research Kolkata, Mohanpur, 741246, India \label{IISERK}}\and\fi
\ifnum\wm>1{Center of Excellence in Space Sciences India, Indian Institute of Science Education and Research Kolkata, 741246, India \label{CESSI}}\and \fi
{School of Physics, Trinity College Dublin, College Green, Dublin 2, D02 PN40, Ireland\label{tcd}}\and
{Jodrell Bank Centre for Astrophysics, Department of Physics and Astronomy, University of Manchester, Manchester M13 9PL, UK\label{jbca}}\and
\ifnum\wm>1{Department of Physics, St. Xavier’s College (Autonomous), Mumbai 400001, India \label{XCM}}\and\fi
{National Astronomical Observatories, Chinese Academy of Sciences, Beijing 100101, P. R. China\label{naoc}}\and
{E.A. Milne Centre for Astrophysics, University of Hull, Cottingham Road, Kingston-upon-Hull, HU6 7RX, UK\label{milne}}\and
{Centre of Excellence for Data Science, Artificial Intelligence and Modelling (DAIM), University of Hull, Cottingham Road, Kingston-upon-Hull, HU6 7RX, UK\label{daim}}\and
\ifnum\wm>1{Department of Physics, BITS Pilani Hyderabad Campus, Hyderabad 500078, Telangana, India \label{BITS}}\and\fi
\ifnum\wm>1{Joint Astronomy Programme, Indian Institute of Science, Bengaluru, Karnataka, 560012, India \label{IISc}}\and\fi
{Arecibo Observatory, HC3 Box 53995, Arecibo, PR 00612, USA\label{arecibo}}\and
{IRFU, CEA, Université Paris-Saclay, F-91191 Gif-sur-Yvette, France \label{irfu}}\and
\ifnum\wm>1{Raman Research Institute India, Bengaluru, Karnataka, 560080, India \label{RRI}}\and\fi
{Institut f\"{u}r Physik und Astronomie, Universit\"{a}t Potsdam, Haus 28, Karl-Liebknecht-Str. 24/25, 14476, Potsdam, Germany\label{uni-potsdam}}\and
\ifnum\wm>1{Department of Physics, IISER Bhopal, Bhopal Bypass Road, Bhauri, Bhopal 462066, Madhya Pradesh, India \label{IISERB}}\and\fi
{Ollscoil na Gaillimhe --- University of Galway, University Road, Galway, H91 TK33, Ireland\label{uog}}\and
\ifnum\wm>1{Center for Gravitation, Cosmology, and Astrophysics, University of Wisconsin-Milwaukee, Milwaukee, WI 53211, USA \label{CGCA}}\and\fi
\ifnum\wm>1{Division of Natural Science, Faculty of Advanced Science and Technology, Kumamoto University, 2-39-1 Kurokami, Kumamoto 860-8555, Japan \label{KU_J1}}\and\fi
\ifnum\wm>1{International Research Organization for Advanced Science and Technology, Kumamoto University, 2-39-1 Kurokami, Kumamoto 860-8555, Japan \label{KU_J2}}\and\fi
{Laboratoire Univers et Th{\'e}ories LUTh, Observatoire de Paris, Universit{\'e} PSL, CNRS, Universit{\'e} de Paris, 92190 Meudon, France\label{luth}}\and
{Florida Space Institute, University of Central Florida, 12354 Research Parkway, Partnership 1 Building, Suite 214, Orlando, 32826-0650, FL, USA\label{FSI}}\and
{Ruhr University Bochum, Faculty of Physics and Astronomy, Astronomical Institute (AIRUB), 44780 Bochum, Germany \label{airub}}\and
{Advanced Institute of Natural Sciences, Beijing Normal University, Zhuhai 519087, China \label{bnuz}}\\
\ifnum\wm>1\\ {\defpta{epta} : The European Pulsar Timing Array}\fi
\ifnum\wm>1\\ {\defpta{inpta} : The Indian Pulsar Timing Array}\fi
}

   \date{TBD}

    \titlerunning{Noise Budget}
\authorrunning{European Pulsar Timing Array DR2}
 
  \abstract
   {}
   {The nanohertz gravitational wave background (GWB) is  expected to be an aggregate signal of an ensemble of gravitational waves emitted predominantly by a large population of coalescing supermassive black hole binaries in the centres of merging galaxies. Pulsar timing arrays, ensembles of extremely stable pulsars at $\sim$ kiloparsec distances precisely monitored for decades, are the most precise experiments capable of detecting this background. However, the subtle imprints that the GWB induces on pulsar timing data are obscured by many sources of noise that occur on various timescales. These must be carefully modelled and mitigated to increase the sensitivity to the background signal.}
   {In this paper, we present a novel technique to estimate the optimal number of frequency coefficients for modelling achromatic and chromatic noise, while selecting the preferred set of noise models to use for each pulsar. We also incorporate a new model to fit for scattering variations in the Bayesian pulsar timing package \texttt{temponest}.  These customised noise models enable a more robust characterisation of single-pulsar noise. We developed a software package based on \texttt{tempo2} to create realistic simulations of European Pulsar Timing Array (EPTA) datasets that allowed us to test the efficacy of our noise modelling algorithms.}
   {Using these techniques, we present an in-depth analysis of the noise properties of 25 millisecond pulsars (MSPs) that form the second data release (DR2) of the EPTA and investigate the effect of incorporating low-frequency data from the Indian Pulsar Timing Array collaboration for a common sample of 10 MSPs. We use two packages, \texttt{enterprise} and \texttt{temponest}, to estimate our noise models and compare them with those reported using EPTA DR1. We find that, while in some pulsars we can successfully disentangle chromatic from achromatic noise owing to the wider frequency coverage in DR2, in others the noise models evolve in a much more complicated way. We also find evidence of long-term scattering variations in PSR\,J1600$-$3053. Through our simulations, we identify intrinsic biases in our current noise analysis techniques and discuss their effect on GWB searches. The analysis and results discussed in this article directly help to improve sensitivity to the GWB signal and are already being used as part of global PTA efforts.}
   {}

    \keywords{(Stars:) pulsars: general --
     (Stars:) pulsars: individual ...  --
     Gravitational waves --
     Methods: statistical}

   \maketitle
%
  
\section{Introduction}
Pulsar timing allows astronomers to track every single rotation of a pulsar. It involves comparing the measured pulse times of arrival (TOAs) with a timing model that contains our current best knowledge of the pulsar's timing properties. 
Millisecond pulsars (MSPs), which are rapidly rotating neutron stars distributed throughout the Galaxy, pulse with sufficient regularity to function as excellent clocks. 
The timing residuals, which are differences between the predicted and estimated arrival times, contain imprints of various astrophysical processes, including the subtle signature induced by the stochastic gravitational wave background (GWB). 

Unlike transient gravitational waves, a GWB will be detected from all directions in the sky. However, a single pulsar is not sufficient to disentangle the contribution of the GWB, as it is affected by many sources of noise with spectral properties similar to those of the GWB. To overcome this, cross-correlations between multiple pulsars are used, as the GWB signal is correlated across multiple sources while the noise (typically intrinsic to the pulsar) is not. Thus, pulsar timing arrays \citep{fb90} extend the concept of pulsar timing to an ensemble of MSPs. However, a GWB is not the only correlated signal amongst an array of pulsars. Imperfections in a common reference clock can cause a monopolar signal \citep{hgc+20}, while unmodeled errors related to solar system ephemerides can induce a dipolar signal \citep{cgl+18}. The GWB is expected to induce a quadrupolar signal, as described by the Hellings-Downs (H-D) curve \citep{hd83}. The latter is considered to be the definitive evidence for the detection of a GWB. It is also important to note that measurable deviations from the H-D curve provide critical insight into the origins of gravitational waves \citep{tmg+15}.

Given the characteristics of the GWB, pulsar timing arrays need to accumulate decades of data to become sensitive to the signal. Three pulsar timing array (PTA) collaborations were formed in the early 2000s, the European Pulsar Timing Array \citep[EPTA;][]{dcl+16}, the North American Nanohertz Observatory for Gravitational Waves \citep[NANOGrav;][]{dfg+13}, and the Parkes Pulsar Timing Array \citep[PPTA;][]{mhb+13}. Since the sensitivity of the GWB relies, among other things, on the number of pulsars and the length of the dataset, the three PTA collaborations came together to form the International Pulsar Timing Array (IPTA).  The Indian Pulsar Timing Array \citep[InPTA;][]{tnr+22} has recently also joined the IPTA. The three PTAs, along with the IPTA, have recently detected a common red signal (CRS) that shares many characteristics with the expected GWB \citep{abb+20,gsr+21,ccg+21,aab+22}. Although these measurements could be the first sign of the GWB emerging from noise, there was no definitive detection of the H-D correlation. The EPTA DR2 with longer timing baselines and more pulsars was created to better constrain the CRS and potentially detect the H-D curve.

Using $\sim$24 years of data and six of the best-timed MSPs, EPTA's detection of the CRS had an amplitude of $2.95^{+0.89}_{-0.72}\times10^{-15}$ for a fixed spectral index ($\gamma$) of $13/3$ (cf. Equation \ref{noise_PSD}) as expected for a GWB originating from a population of supermassive black hole binaries \citep{ccg+21}. Further investigations of the spectral properties of individual pulsars by \citet{cbp+22} showed a clear need for pulsar-specific noise models and frequency binning. Furthermore, \citet{gsr+21,gts+22} stated that it is likely that at least part of the CRS can arise from a superposition of independent RN processes in individual MSPs. 
With simulated datasets, \citet{zhs+22} demonstrated how spurious evidence for CRS arises based on the choice of priors. Thus, robust modelling of the noise within each MSP dataset is critical given the possibility that the CRS may be a subset of the GWB signal.  

In addition to radio PTAs, an independent limit on the GWB was placed using gamma ray observations of MSPs \citep{fermi_pta_2022}. This not only provides an independent handle on the properties of the GWB but also, since gamma rays are immune to the effects of the ionised interstellar medium (IISM), it offers the possibility of testing the efficacy of current chromatic models through joint radio and gamma-ray analysis. Therefore, it is natural for such analyses to be integrated into future IPTA data releases. 

In this paper, we will focus on the long-term timing noise budget of pulsars in the second data release of the EPTA (hereafter, DR2full). This contains $\sim$ 25 years of data for 25 of the most precisely timed MSPs in the EPTA DR1. This new dataset will add significant sensitivity at the lowest Fourier frequencies of the GWB, owing to its long time span, and increased sensitivity at the higher Fourier frequencies due to improved observing cadence and an effective doubling of the observing bandwidth. We also combine low radio frequency pulsar timing data from the InPTA for 10 MSPs that are common to the EPTA DR2 (hereafter, DR2full+), to investigate the effects of the IISM and their impact on the total noise budget. For details on the EPTA and InPTA datasets, observing systems, and pulsar timing parameters, refer to \citet{wm1}.

The paper is structured as follows. In Section \ref{noisebudget}, we provide an overview of the noise budget, the descriptions of techniques used to estimate optimal Fourier frequency bins, and the selection of favoured noise models. In Section \ref{results}, we present the estimated noise budget for the EPTA and InPTA datasets and highlight interesting cases. In Section \ref{interpretation}, we interpret our results, compare them with previous analyses, and discuss implications for GWB searches. In Section \ref{noisevalidation}, we discuss the efficacy of our noise models through additional tests, simulations, and cross-validation with independent software packages. In Section \ref{conclusions}, we present conclusions and provide potential future trajectories to explore, especially by focussing on IPTA datasets. The customised noise models reported in this paper are applied directly to the EPTA's search for spatially correlated GWB signals \citep{wm3}.

\section{Noise Modelling Techniques and Noise Budget}
\label{noisebudget}

The assumption that TOAs are only composed of a deterministic timing model plus time-uncorrelated instrumental noise is generally untrue in pulsar timing datasets. Stochastic processes, ranging from intrinsic spin noise in the neutron star to chromatic variations resulting from the IISM, contribute additional time-correlated noise in the observed TOAs \citep{Cordes_Shannon_2010}. Unless this noise is accounted for, it can bias timing model parameter estimation and reduce the sensitivity to the GWB. 
The typical method to account for this noise is to solve for both the pulsar timing model and a stochastic Gaussian-process noise model simultaneously \citep{temponest_physrevd,rutger2014}.
This is achieved by the use of Bayesian parameter estimation routines to find the posterior probability distribution of the noise model hyperparameters and the pulsar timing model parameters.
In practice, significant computation time can be saved by analytically marginalising over most or all of the pulsar timing model parameters.
This is achieved by reusing the linearized approximation to the timing model adopted in traditional pulsar timing software such as \texttt{tempo2}, which significantly reduces the parameter space to be sampled in the Bayesian analysis.
The preparation of the parameters of the pulsar timing model, including the choice of parameters included in the fit, is discussed in \citet{wm1}.

\subsection{Description of the noise models}
\label{noise_models}
For the work discussed in this paper, we focus primarily on the standard noise models used widely within the IPTA (\citet{temponest_physrevd}, \citet{lhf+14} and references therein). In brief, we model the excess noise in the pulsar as a combination of Gaussian processes with two main forms, either correlated in time (i.e. `red' noise) or uncorrelated in time (i.e. `white' noise). Red noise usually dominates on timescales of years to decades and incorporates signals with power spectral densities similar to the GWB. These processes include both chromatic noise, which depends on observing frequency, and achromatic noise, which is independent of observing frequency. We generally consider one achromatic red timing noise and two chromatic processes, variation in the dispersion measure (DM) to the pulsar, and variation in the scattering timescale for the pulsar. White noise reflects unmodeled instrumental errors and intrinsic pulse jitter \citep{lvk+11,lkl+12,ap_jitter} in the arrival times.

\subsubsection{Time-correlated noise}
Pulsar timing noise represents stochastic irregularities in pulsar rotation.
Persistent temporally correlated noise that manifests equally across the radio frequency band of the instrument is referred to as \emph{spin noise}, an achromatic noise process.
It is typically modelled as a wide-sense stationary stochastic signal, i.e. a process with zero mean and a covariance function that depends only on the absolute time lag between two points.
Timing noise is present across the pulsar population and its amplitude seems to vary as a function of the pulsar spin-down rate \citep{sc10}. Although the origins of spin noise are not unanimously agreed upon, it is typically well-modelled with a power-law spectrum.
Power-law behaviour is expected if spin noise originates due to interactions between the neutron star crust and its superfluid core \citep{jones_1990}, although observations of several canonical pulsars (i.e. non-MSPs) show a quasi-periodic behaviour in timing noise properties \citep{lhk+10}, or spectral turnovers \citep{Parthasarathy19} that may warrant additional terms beyond a single power law. However, the relatively small amplitudes of spin noise seen in MSPs is found to be relatively well-modelled by a power-law process \citep{gzt20}.

The IISM introduces a wide range of chromatic noise processes into the TOAs that are dependent on the observing frequency. These include dispersive delays, scintillation, and pulse profile broadening due to multipath propagation \citep{Cordes16,Shannon17,Donner20}. Dispersive delays can become measurable on timescales of days to weeks and scale with the observing radio frequency as $\nu^{-2}$. 
DM variations are one of the biggest sources of unmodeled error, and although the wider bandwidths and higher cadence observations of recent PTA datasets significantly improves the ability to model the DM variations and separate them from achromatic signals, the inhomogeneity of the datasets means that it can be difficult to separate chromatic and achromatic noise on the longest timescales.
The best approach to account for DM variability depends on the underlying dataset. For EPTA, we currently favour using a stochastic power-law model for the DM variations, which assumes that there is a stationary smooth process that determines the DM. This allows for observations separated in both time and observing frequency to be used for constraining the DM. The theoretical expectation is that DM variations are caused by Kolmogorov turbulence in the IISM, and hence will have a power law index of $\gamma_\mathrm{DM} = 8/3$ \citep{fc90}. Alternative models (such as the use of \texttt{DMX} parameters \citep{Alam21}), which typically make independent measurements of DM over discrete time intervals, avoid the assumption of smooth variation and stationary process, but require near-simultaneous observations at a wide range of frequencies to be effective. Additionally, there is a variable contribution to the DM from the solar wind. A study of the time-variable solar wind is part of an ongoing EPTA project (Ni\c{t}u et al., in preparation). For this work, we use a fixed value of 7.9 \,cm$^{-3}$ \citep{mca+19} for all pulsars, except for PSR\,J1022+1001 where we fit a constant solar wind amplitude \footnote{The solar wind amplitude is measured at 1 au} \citep[see also ][]{wm1}.

After DM variations, the second most prominent effect of the interstellar medium at typical observing frequencies is the variation of the pulsar's scattering timescale. This scales as $\nu^{-4}$ and therefore can be separated from DM given enough coverage in observing frequency. Variation in scattering timescale is modelled as a power law in the few cases where it is important for the pulsars in our current dataset.

For each of our time-correlated noise processes, we model the noise of a process with 
chromatic index $\alpha$ for a TOA at time $t$ and observing frequency $\nu$ ($\nu_\mathrm{ref}$ is the reference frequency set at 1400 MHz) with a 
Fourier basis of $N_\mathrm{coef}$ coefficients as,
\begin{equation}
    y(t) = \sum\limits_{j=1}^{N_\mathrm{coef}} Y_j \left(a_j \cos{(j\omega t)} + b_j 
    \sin{(j\omega t)}\right)\left(\frac{\nu}{\nu_\mathrm{ref}}\right)^{-{\alpha}},
    \label{noise_equation}
\end{equation}
where $\omega = 2\pi/T_\mathrm{span}$ for $T_\mathrm{span}$ typically chosen to be the total observing time span, $a_j$ and $b_j$ are fit parameters with a Gaussian prior $\mathcal{N}(0,1)$ and $Y_j$ determined by the noise model hyperparameters $A$ and $\gamma$. Specifically, we define $Y_j$ as
\begin{equation}
    Y_j = \sqrt{\frac{A^2}{K_\mathrm{scale}}\frac{\mathrm{s_\mathrm{yr}}^3}{T_\mathrm{span}}\left(\frac{f_j}{f_\mathrm{yr}}\right)^{-\gamma}},
    \label{noise_PSD}
\end{equation}
where $f_\mathrm{yr} = 1 \mathrm{yr}^{-1}$,  $\mathrm{s_\mathrm{yr}} = 31557600\,\mathrm{s\,yr}^{-1}$ converts years to seconds with $T_\mathrm{span}$ in seconds, and $K_\mathrm{scale}$ is a scale that can adjust the units of $A$ appropriately for the given noise process.
In practise, the parameters $a_j$ and $b_j$ are analytically marginalised following the method described in \citet{lhf+14} and implemented in \texttt{temponest} and \texttt{enterprise}.

The choice of constants for each process, red noise, DM noise and scattering variations, and the hyperparameter priors are given in Table \ref{model_priors}. Achromatic red noise is scaled so that $A_\mathrm{red}$ is the equivalent GWB amplitude in $\mathrm{yr}^{3/2}$, the DM variations are given in \texttt{temponest} units of $\mathrm{cm}^{-3}\mathrm{pc}\,\mathrm{yr}^{3/2}\mathrm{s}^{-1}$, and the
scattering variations are given as the equivalent amplitude of the red noise (in $\mathrm{yr}^{3/2}$) at $1400\,$MHz. These units are chosen to match previous publications and can be used directly in \texttt{tempo2}. 

\begin{table} 
    \centering
    \caption{Noise model constants and hyperparameter priors. All priors are uniform between the specified bounds. The term $k_\mathrm{DM} = 2.41\times10^{-4} \mathrm{cm^{-3}pc\,MHz^2s^{-1}}$ is the DM constant.}
    \label{model_priors}
    \begin{tabular}{c|c|c|c}
    \toprule
      Parameter & Red    &  DM &   Scatter \\
            \midrule             
        $K_\mathrm{scale}$ & $12\pi^2$ & $k_\mathrm{DM}^2$ & $12\pi^2$ \\
        $\alpha$ & 0 & 2 & 4\\
        $\mathrm{Prior}(\log_{10}(A))$ &$\mathcal{U}(-18,-10)$ & $\mathcal{U}(-18,-10)$ & $\mathcal{U}(-18,-10)$\\
        $\mathrm{Prior}(\gamma)$ & $\mathcal{U}(0,7)$ & $\mathcal{U}(0,7)$ & $\mathcal{U}(0,7)$\\
        \bottomrule
    \end{tabular}

\end{table}

\subsubsection{White noise}
Temporally-uncorrelated white noise in pulsar timing residuals needs to be modelled to effectively estimate the precision of pulsar timing parameters. For this work, we include the widely adopted parameters \footnote{However, some PTA collaborations use different notations} \textsc{efac} and \textsc{equad} where the diagonals of the noise covariance matrix are scaled by
\begin{equation}
\sigma_\mathrm{scaled}^2 = \mathrm{EFAC}^2 \times \sigma^2_\mathrm{original} + \mathrm{EQUAD}^2,
\label{efacequad}
\end{equation} 

where $\sigma_{\rm{original}}$ is the measurement uncertainty of a TOA due to template-fitting errors. These parameters are applied for every observing backend and are tagged with the \texttt{-group} flag in the EPTA dataset. We use uniform priors on EFAC (0.1 to 5) and $\log_{10}(\mathrm{EQUAD})$ ($-9$ to $-5$).

\subsubsection{Exponential dips}
\label{expdip}
The exponential dips are signals manifesting as a sudden radio frequency-dependent advance of pulse arrival times and are estimated to impact the measurements of time-correlated signals \citep{hst+20}. Such events have been reported three times for PSR J1713+0747 at MJDs $\sim 54757 (2008)$ \citep{cks2015,zsd2015,dcl+16}, $\sim 57510 (2016)$ \citep{leg+18,gsr+21,cbp+22} and $\sim 59320 (2021)$ \citep{xhb2021,abb2021,ssj2021}. Only the first two events are part of DR2full. This effect can be explained by a sudden drop in the density of the IISM electron column along the line of sight to the pulsar and would induce an effect on the timing residuals with a chromatic index of $2$. It could also be caused by a magnetospheric process inducing a profile shape change with a different chromatic index than for a DM event \citep{slk1016,gsr+21}.

The model commonly used in \texttt{enterprise} describes the exponential dips as a deterministic signal with a waveform $\vec{d_{\mathrm{exp}}}$ expressed for any observing time $t_i > t_0$ and radio frequency $\nu_k$ as
\begin{equation}
    \vec{d_{\mathrm{exp}}} (t_i, \nu_k \ ; \vec{\boldsymbol \theta}_{\mathrm{exp}}) =
        A \ \left( \frac{\nu_k}{\nu_{\mathrm{ref}}} \right)^{-\alpha} \ \mathrm{exp} \left( - \frac{t_i - t_0}{\tau} \right),
\end{equation}
with the reference frequency $v_{\mathrm{ref}}$ set at $1.4$ GHz and the parameters $\vec{\boldsymbol \theta}_{\mathrm{exp}}$ being the amplitude ($A$; in residual units), the relaxation time ($\tau$; in days), the initial epoch ($t_0$; in MJD) and the chromatic index ($\alpha$). The prior probability distributions are described in Table \ref{expdip_prior}. For the analyses in this work and similarly to \citet{cbp+22}, we fix the exponential dips $\alpha$ of the first and second events, respectively, at $4$ and $1$, chosen from their marginalised posteriors evaluated a priori.

\begin{table} 
    \centering
    \caption{Parameter priors for the two exponential dips included in the noise models for PSR J1713+0747.}
    \label{expdip_prior}
    \begin{tabular}{c|c}
    \toprule
      Parameter & Prior \\
            \midrule             
        A [s] & $\mathrm{log}_{10} \mathcal{U} (10^{-10}, 10^{-2})$ \\
        $\tau$ [day] & $\mathrm{log_{10} {\mathcal{U}}(1, 10^{2.5}})$ \\
        $t_0$ (1st event) [MJD] & $\mathcal{U}(54650, 54850)$\\
        $t_0$ (2nd event) [MJD] & $\mathcal{U}(57490, 57530)$\\
        $\alpha$ (1st event) & 4 (fixed value) \\
        $\alpha$ (2nd event) & 1 (fixed value) \\
        \bottomrule
    \end{tabular}

\end{table}

\subsection{Bayesian inference for pulsar timing noise analysis}
We use a Bayesian approach (e.g. \cite{Sivia2006}) for the parameter estimation and model selection. Given the measured timing residuals $\vec{\delta t}$, the parameters $\vec{\boldsymbol \theta}_i$ of a chosen model $\mathcal{M}_i$ are considered random variables with a probability density function (\emph{parameter posteriors}) described with the Bayes theorem as follows.
\begin{equation}
    p(\vec{\boldsymbol \theta}_i \ | \ \vec{\delta t}, \mathcal{M}_i) = \frac{p( \vec{\delta t} \ | \ \vec{\boldsymbol \theta}_i, \mathcal{M}_i) \ p(\vec{\boldsymbol \theta}_i \ | \ \mathcal{M}_i)}{p( \vec{\delta t} \ | \ \mathcal{M}_i)},
    \label{BayesTh}
\end{equation}
where $p(\vec{\boldsymbol \theta}_i \ | \ \mathcal{M}_i)$, $p( \vec{\delta t} \ | \ \vec{\boldsymbol \theta}_i, \mathcal{M}_i)$ and $p( \vec{\delta t} \ | \ \mathcal{M}_i)$ are the \emph{parameter prior}, \emph{likelihood} and \emph{marginal likelihood} (or evidence), respectively. The prior distributions for the noise parameters are described in Section \ref{noise_models} and those for the timing model parameters are shown in \citep{wm1}. 

We use a Gaussian likelihood \citep{rutger2015,ac2021}, where the deterministic signal waveforms are directly reduced to the timing residuals and the stochastic time-correlated and uncorrelated components are included in the time-domain covariance matrix. The evidence corresponds to the marginalised likelihood, that is, the integral of the likelihood over the whole parameter space. This term is particularly useful for model selection, as described below.

Let us now rewrite the Bayes theorem to express the probability density function of a model $\mathcal{M}_i$ given the set of timing residuals $\vec{\delta t}$
\begin{equation}
    p(\mathcal{M}_i \ | \ \vec{\delta t}) = \frac{p( \vec{\delta t} \ | \ \mathcal{M}_i) \ p(\mathcal{M}_i)}{p(\vec{\delta t})},
\end{equation}
where $p(\mathcal{M}_i)$ is the prior of model $\mathcal{M}_i$, $p( \vec{\delta t} \ | \ \mathcal{M}_i)$ is the evidence that also appears in equation \ref{BayesTh} and $p(\vec{\delta t})$ is the probability of observing the timing residuals marginalised over all models. The odds ratio, that is, the probability ratio between two models $\mathcal{M}_1$ and $\mathcal{M}_2$ given the same data is defined as
\begin{equation}
    \frac{p(\mathcal{M}_2 \ | \ \vec{\delta t})}{p(\mathcal{M}_1 \ | \ \vec{\delta t})} = \frac{p( \vec{\delta t} \ | \ \mathcal{M}_2)}{p( \vec{\delta t} \ | \ \mathcal{M}_1)} \ \frac{p(\mathcal{M}_2)}{p(\mathcal{M}_1)}.
    \label{BayesTh2}
\end{equation}
Considering equal prior probability between all models (as in this work), equation \ref{BayesTh2} is reduced to the ratio of evidences, also termed the \emph{Bayes factor},
\begin{equation}
    \mathcal{B}^{\mathcal{M}_2}_{\mathcal{M}_1} = \frac{p( \vec{\delta t} \ | \ \mathcal{M}_2)}{p( \vec{\delta t} \ | \ \mathcal{M}_1)}.
\end{equation}
If $\mathcal{B}^{\mathcal{M}_2}_{\mathcal{M}_1} > 1$, the model $\mathcal{M}_2$ is preferred over $\mathcal{M}_1$ given the measured timing residuals $\delta t$ considered as the data. We use the scale proposed in \citet{kass_rafferty} to interpret and make decisions about the selection of the model and include an additional noise component in a simpler model only if the related Bayes factor is equal to or greater than $150$. In this work, most of the Bayes factors are evaluated following a product-space sampling approach \citep{carlin_chib_1995,Hee_2015,tvs20}, also referred to as the ``HyperModel'' method, where an additional hyperparameter is added to switch between two or more models. The Bayes factor between the two models becomes equivalent to the ratio of the number of samples between the models.

\subsection{Noise modelling packages and samplers}
For the single-pulsar noise analysis results reported in this paper, we use two Bayesian pulsar timing analysis packages, \texttt{temponest} \citep{lhf+14} and \texttt{enterprise}\footnote{https://gitlab.in2p3.fr/epta/enterprise/-/tree/master/enterprise} \citep{enterprise}.
Both implement the noise models as described in Section \ref{noise_models}. Although both codes have been tested with a range of samplers, for this work, the sampling for \texttt{temponest} is performed using Multinest \citep{multinest1} and for \texttt{enterprise} is performed using PTMCMCSampler \citep{ptmcmc}. We have made several optimisations to \texttt{temponest}\footnote{https://github.com/aparthas3112/Temponest}, including adding a scattering delay variation model.

\subsection{Choosing the optimal number of Fourier Coefficients}
\label{nbins}
We model the red noise, the DM variations and the scattering variations as stationary Gaussian processes, following the Fourier-sum approach described in \citet{rutger2014}, with a discrete and finite set of sine/cosine basis functions and a power-law power spectral density (PSD). The set of frequencies for each Gaussian process is chosen to be linearly distributed as $1/T_\mathrm{span}, 2/T_\mathrm{span}, ..., N_\mathrm{coef}/T_\mathrm{span}$. Most PTA analyses in the past used the same number of frequency bins $N_{\mathrm{coef}}$ for a given signal for all pulsars. However, different types of covariance expansions have been discussed in \citet{rutger2015}, and methods to select a customised number of frequency bins have been proposed, either from the transitional frequency of a ``red noise - white noise'' broken power-law PSD \citep{ccg+21}, or from selecting the preferred value from a model selection among chosen sets of frequency bins \citep{ac2021}. In this work, we propose a novel method to select the number of frequency components for Gaussian processes by setting $N_{\mathrm{coef}}$ as a free parameter in the noise model, and thus evaluating a marginalised posterior distribution $p(N_{\mathrm{coef}} | \ \vec{\delta t})$. This approach extends the method performed in \cite{ac2021} by enabling tests for any possible value in a selected range of frequencies without having to evaluate a Bayes factor between all possible models. As the prior is the same between two Gaussian processes with a different number of frequency bins, this method is equivalent to a Bayes factor evaluation from a product-space approach.

We decide to set a non-informative prior for the discrete parameter $N_{\mathrm{coef}}$. The method implemented consists of including a real hyperparameter $\tilde{N}_{\mathrm{coef}}$ with a uniform prior \mbox{$\mathcal{U} \left( [ \tilde{N}_{\mathrm{min}}, \tilde{N}_{\mathrm{max}}+1 [ \right)$}, and setting the number of frequency bins $N_{\mathrm{coef}}$ as the integer floor value $\left\lfloor \tilde{N}_{\mathrm{coef}} \right\rfloor$, thus ensuring an equal probability between all frequencies inside the prior range. The prior limits for the three stochastic signals considered in this work (red noise, DM, and scattering delay variations) are $\tilde{N}_{\mathrm{min}}=10$ and $\tilde{N}_{\mathrm{max}}=150$. The former is empirically chosen for statistical reasons. The maximum value of $150$ allows testing for frequencies up to $\sim 33 \ \mathrm{days}^{-1}$ and $\sim 60 \ \mathrm{days}^{-1}$ for pulsars with the shortest and longest baselines, respectively, where white noise is expected to dominate over other signals. The value of the upper edge prior for the truncated dataset (DR2new) (cf. Section \ref{results}) is $\tilde{N}_{\mathrm{max}} = 100$ to cover frequencies up to $37 - 33 \ \mathrm{days}^{-1}$ for all pulsars.

In case of an inconclusive (i.e. flat) posterior distribution, we select a minimal value and perform an additional Bayes factor evaluation between the maximum posterior value and the selected minimal value as a performance check. By minimizing the number of Fourier modes, we aim at selecting only the low-frequency noise expected to follow a power-law PSD, and reduce the effects of excess noise at high frequencies in the white noise dominated range. Furthermore, this step allows to reduce the computational cost of the likelihood which becomes crucial for a gravitational-wave analysis where the Gaussian process hyperparameters for all the pulsars in the array are simultaneously sampled.

\subsection{Model selection and parameter estimation methodology}
\label{modelselection}

For each pulsar, we perform a Bayesian model selection to evaluate the most favoured combination among a chosen list of time-correlated noise components (cf. Section \ref{noise_models}). The selection between two models is based on a Bayes factor evaluation performed with \texttt{enterprise} using the HyperModel method. For these analyses, we marginalise across all timing model parameters and fit the remaining parameters. For PSR J1713+0747, we also include the exponential dips described in Section \ref{expdip}. After obtaining the customised noise models, a full Bayesian analysis is performed with \texttt{temponest} that fits all the timing and noise model parameters. The results of these analyses are reported in Section \ref{results}. The procedure used to obtain the customised noise models is described in the following part of this section.

The six candidate noise models include any combination of achromatic red noise (RN), DM variations (DM), and scattering delay variations (SV). These are \textsc{none} (no time-correlated noise), \textsc{rn}, \textsc{dm}, \textsc{rn+dm}, \textsc{dm+sv}, \textsc{rn+dm+sv} (as reported in Tables \ref{eptatable1} and \ref{eptatable2}. In this work, we do not consider models including SV but without DM, as they are physically unlikely at the observed radio frequencies in our dataset. We first evaluate the number of frequency bins that are the most preferred for each candidate noise model following the method described in Section \ref{nbins}. In the second step, we select the most favoured model among the six candidates with a Bayes factor evaluation. Following the principle of Occam's razor, we select the simplest model (i.e. the model with the lowest number of Gaussian process signals) in case of inconclusive results.

After obtaining the final customised models, we perform additional checks to consolidate the robustness of the results described in Section \ref{noisevalidation}. This step has been particularly useful in revisiting the noise model for PSR J1022+1011, which exhibited unmodelled achromatic red noise in the low frequencies of the power spectrum. A deeper analysis and a more advanced noise model would likely be required for this pulsar. Therefore, we stick to a standard version of the noise model for this pulsar, including RN and DM with $30$ and $100$ PSD frequencies, respectively.

\section{Estimated noise budget for the EPTA DR2 pulsars}
\label{results}
Here we present the results of the noise model selection and parameter estimation for two EPTA datasets (to assist the reader, Table \ref{datasetreference} provides a summary of the various datasets used in this paper),  
\begin{itemize}
    \item The first is DR2full, containing data from the EPTA DR1 and the new instrumentation in DR2.
    \item The second dataset consists only of the EPTA DR2 data from the new instrumentation (starting from $\sim$ MJD 55611; hereafter, DR2new). 
    \item The third is DR2full+ which contains the data described as DR2full along with additional low-frequency data from the InPTA collaboration for 10 common MSPs. 
\end{itemize}

\begin{table}
    \centering
    \caption{Abbreviations for the different datasets used in this paper. The InPTA dataset extends the timing baseline of the EPTA dataset from MJD~59385 to 59644. It also provides consistent multi-frequency coverage for a period of a few years, down to the low-end ($\sim300$\,MHz) of the frequency range whose coverage is very sparse with EPTA data only.}
    \begin{tabular}{l|l}
    \hline
        \textbf{Dataset} & \textbf{Referenced as} \\\hline
        EPTA data with 25 MSPs & DR2full \\ 
        EPTA data with only new backends & DR2new \\ 
        DR2full + InPTA data & DR2full+ \\ \hline
    \end{tabular}
    \label{datasetreference}
\end{table}

In principle, the combined EPTA+InPTA dataset (DR2full+) is much more sensitive for measuring the pulsar noise parameters as well as the GWB, with a total time span of 15--25 years for most pulsars. However, around half of these data are from instruments that were designed prior to the start of the PTA experiment, and in some cases lacked coherent dedispersion or were otherwise limited in time resolution on the fastest pulsars.
Furthermore, in some cases the early data do not have good coverage in observing frequency, making it difficult to disentangle chromatic and achromatic noise processes.
Hence, although the DR2new is only 10 years long, it may contain fewer systematics and degeneracies in the noise models. Indeed, as shown in \citet{wm3}, DR2new appears to be favourable in sensitivity to the GWB, which may suggest that the noise modelling is insufficient for early EPTA data, or that the characteristics of the pulsar noise or GWB signal are varying with time. Further investigation of the time-stationary properties of noise in each pulsar is briefly discussed in Section \ref{noisecompare} and should be a priority for future IPTA data analysis. 

In Table \ref{eptatable1} we present the noise budget for 25 MSPs in DR2full+, while Table \ref{eptatable2} shows the same results when applied to DR2new. For each pulsar, we report the properties of the preferred noise model estimated from the method described in Section \ref{modelselection} and discuss the interpretations of the results in Section \ref{noisecompare}.

\begin{table*}
\caption{Favoured models listed for the 25 pulsars using both EPTA and InPTA data (for 10 common MSPs; DR2full+) along with estimated values for chromatic and achromatic noise models. The second column lists the PTAs that contributed to the dataset for each pulsar. The third column reports the favoured model. Columns 4 to 9 report the estimated number of coefficients, amplitude, and slope (medians with $95\%$ confidence intervals) for the achromatic and chromatic noise processes, respectively. The last column reports the total time span for each of the pulsars.}
\label{eptatable1}
\begin{tabular*}{\textwidth}[c]{@{\extracolsep{\fill}} l|l|l|ccc|ccc|c}
\toprule
Pulsar & 
PTA &
Favoured  & 
  \multicolumn{3}{c|}{Red noise} &
  \multicolumn{3}{c|}{DM noise} &
  \multicolumn{1}{c}{Time span} \\
  
  \cmidrule(lr){4-6} \cmidrule(lr){7-9} &
  \multicolumn{1}{c}{} & 
  \multicolumn{1}{c}{Models} &
  \multicolumn{1}{c}{$N_{\mathrm{coef}}$} &
  \multicolumn{1}{c}{$A$} &
  \multicolumn{1}{c}{$\gamma$} &
  \multicolumn{1}{c}{$N_{\mathrm{coef}}$} &
  \multicolumn{1}{c}{$A$} &
  \multicolumn{1}{c}{$\gamma$} &
  \multicolumn{1}{c}{yr} \\ \midrule
  
J0030+0451 & EPTA & RN & $10$ & ${-14.93}^{+0.83}_{-1.1}$ & ${5.49}^{+1.93}_{-1.56}$ & X & X & X & 21.96 \\[3pt]
J0613$-$0200 & EPTA+InPTA & RN+DM & $10$ & ${-14.99}^{+0.94}_{-1.24}$ & ${5.34}^{+2.06}_{-1.6}$ & $129$ & ${-11.58}^{+0.06}_{-0.06}$ & ${1.34}^{+0.28}_{-0.26}$ & 23.83 \\[3pt]
J0751+1807 & EPTA+InPTA & DM & X & X & X & $115$ & ${-11.72}^{+0.2}_{-0.2}$ & ${2.69}^{+0.51}_{-0.49}$ & 25.12 \\[3pt]
J0900$-$3144 & EPTA & RN+DM & $135$ & ${-12.76}^{+0.09}_{-0.08}$ & ${1.06}^{+0.28}_{-0.27}$ & $150$ & ${-11.94}^{+0.67}_{-0.87}$ & ${3.89}^{+2.12}_{-1.79}$ & 13.64 \\[3pt]
J1012+5307 & EPTA+InPTA & RN+DM & $149$ & ${-13.03}^{+0.05}_{-0.04}$ & ${1.21}^{+0.17}_{-0.17}$ & $47$ & ${-11.95}^{+0.11}_{-0.12}$ & ${1.74}^{+0.39}_{-0.37}$ & 24.61 \\[3pt]
J1022+1001 & EPTA+InPTA & RN+DM & $30$ & ${-13.8}^{+0.51}_{-0.99}$ & ${3.01}^{+1.55}_{-0.97}$ & $100$ & ${-11.46}^{+0.09}_{-0.08}$ & ${0.14}^{+0.26}_{-0.13}$ & 25.37 \\[3pt]
J1024$-$0719 & EPTA & DM & X & X & X & $34$ & ${-11.82}^{+0.18}_{-0.21}$ & ${2.46}^{+0.87}_{-0.66}$ & 23.14 \\[3pt]
J1455$-$3330 & EPTA & RN & $49$ & ${-13.26}^{+0.28}_{-0.49}$ & ${2.21}^{+1.35}_{-1.04}$ & X & X & X & 15.72 \\[3pt]
J1600$-$3053 & EPTA+InPTA & RN+DM & $21$ & ${-14.05}^{+0.49}_{-0.89}$ & ${2.86}^{+1.99}_{-1.24}$ & $148$ & ${-11.46}^{+0.04}_{-0.04}$ & ${1.99}^{+0.12}_{-0.12}$ & 15.42 \\[3pt]
J1640+2224 & EPTA & DM & X & X & X & $145$ & ${-11.66}^{+0.14}_{-0.13}$ & ${0.48}^{+0.49}_{-0.4}$ & 24.44 \\[3pt]
J1713+0747 & EPTA+InPTA & RN+DM & $12$ & ${-14.19}^{+0.27}_{-0.29}$ & ${3.28}^{+0.66}_{-0.63}$ & $148$ & ${-11.86}^{+0.05}_{-0.04}$ & ${1.59}^{+0.19}_{-0.19}$ & 24.5 \\[3pt]
J1730$-$2304 & EPTA & DM & X & X & X & $10$ & ${-11.56}^{+0.55}_{-0.57}$ & ${2.22}^{+1.56}_{-1.45}$ & 16.1 \\[3pt]
J1738+0333 & EPTA & RN & $11$ & ${-12.93}^{+0.36}_{-0.4}$ & ${2.14}^{+1.31}_{-1.2}$ & X & X & X & 14.12 \\[3pt]
J1744$-$1134 & EPTA+InPTA & RN+DM & $10$ & ${-14.12}^{+0.41}_{-0.72}$ & ${3.45}^{+1.19}_{-0.75}$ & $150$ & ${-11.82}^{+0.1}_{-0.07}$ & ${0.26}^{+0.37}_{-0.23}$ & 25.14 \\[3pt]
J1751$-$2857 & EPTA & DM & X & X & X & $41$ & ${-11.08}^{+0.22}_{-0.33}$ & ${2.13}^{+0.99}_{-0.7}$ & 14.69 \\[3pt]
J1801$-$1417 & EPTA & DM & X & X & X & $14$ & ${-10.73}^{+0.27}_{-0.26}$ & ${1.68}^{+1.16}_{-1.06}$ & 13.71 \\[3pt]
J1804$-$2717 & EPTA & DM & X & X & X & $38$ & ${-11.19}^{+0.18}_{-0.83}$ & ${0.78}^{+2.95}_{-0.71}$ & 14.73 \\[3pt]
J1843$-$1113 & EPTA & DM & X & X & X & $73$ & ${-11.03}^{+0.08}_{-0.08}$ & ${2.07}^{+0.36}_{-0.31}$ & 16.8 \\[3pt]
J1857+0943 & EPTA+InPTA & DM & X & X & X & $11$ & ${-11.86}^{+0.27}_{-0.28}$ & ${2.88}^{+0.66}_{-0.62}$ & 25.11 \\[3pt]
J1909$-$3744 & EPTA+InPTA & RN+DM & $20$ & ${-14.89}^{+0.78}_{-0.85}$ & ${4.77}^{+1.96}_{-1.79}$ & $150$ & ${-11.85}^{+0.05}_{-0.05}$ & ${1.31}^{+0.16}_{-0.15}$ & 17.14 \\[3pt]
J1910+1256 & EPTA & DM & X & X & X & $10$ & ${-11.71}^{+0.66}_{-0.84}$ & ${2.98}^{+2.38}_{-1.87}$ & 15.21 \\[3pt]
J1911+1347 & EPTA & DM & X & X & X & $10$ & ${-11.98}^{+0.39}_{-0.47}$ & ${3.06}^{+1.36}_{-1.06}$ & 14.2 \\[3pt]
J1918$-$0642 & EPTA & DM & X & X & X & $138$ & ${-12.09}^{+0.4}_{-0.44}$ & ${3.49}^{+1.13}_{-1.06}$ & 19.71 \\[3pt]
J2124$-$3358 & EPTA+InPTA & DM & X & X & X & $18$ & ${-11.77}^{+0.34}_{-0.39}$ & ${2.07}^{+1.09}_{-0.98}$ & 17.15 \\[3pt]
J2322+2057 & EPTA & NONE & X & X & X & X & X & X & 14.68 \\[3pt]
  
\bottomrule
\end{tabular*}%
\end{table*}%

\begin{table*}

\caption{The same as in Table \ref{eptatable1}, but for DR2new. Pulsars for which the preferred noise models have changed from Table \ref{eptatable1} are highlighted.}
\label{eptatable2}
\begin{tabular*}{\textwidth}[c]{@{\extracolsep{\fill}} l|l|ccc|ccc|c}
\toprule
Pulsar & 
Favoured Models & 
  \multicolumn{3}{c|}{Red noise} &
  \multicolumn{3}{c|}{DM noise} &
  \multicolumn{1}{c}{Time span} \\
  
  \cmidrule(lr){3-5} \cmidrule(lr){6-8} &
  \multicolumn{1}{c}{} &
  \multicolumn{1}{c}{$N_{\mathrm{coef}}$} &
  \multicolumn{1}{c}{$A$} &
  \multicolumn{1}{c}{$\gamma$} &
  \multicolumn{1}{c}{$N_{\mathrm{coef}}$} &
  \multicolumn{1}{c}{$A$} &
  \multicolumn{1}{c}{$\gamma$} &
  \multicolumn{1}{c}{yr} \\ \midrule

J0030+0451 & RN &  10 & ${-14.72}^{+1.26}_{-1.29}$ & ${4.6}^{+2.27}_{-3.36}$ & X & X & X & 9.78 \\[3pt]
J0613$-$0200 & \textbf{DM} &  X & X & X & 11 & ${-11.79}^{+0.34}_{-1.0}$ & ${1.14}^{+3.23}_{-1.08}$ & 10.07 \\[3pt]
J0751+1807 & DM &  X & X & X & 92 & ${-11.8}^{+0.32}_{-0.82}$ & ${2.9}^{+2.31}_{-1.04}$ & 10.08 \\[3pt]
J0900$-$3144 & RN+DM & 99 & ${-12.72}^{+0.12}_{-0.13}$ & ${1.19}^{+0.44}_{-0.4}$ & 100 & ${-11.81}^{+0.73}_{-1.22}$ & ${3.78}^{+2.91}_{-2.18}$ & 9.86 \\[3pt]
J1012+5307 & RN+DM & 92 & ${-12.99}^{+0.12}_{-0.1}$ & ${1.25}^{+0.37}_{-0.34}$ & 16 & ${-12.66}^{+0.78}_{-1.55}$ & ${3.93}^{+2.86}_{-2.53}$ & 10.08 \\[3pt]
J1022+1001 & RN+DM & 30 & ${-15.42}^{+1.9}_{-2.44}$ & ${3.79}^{+3.04}_{-3.56}$ & 100 & ${-11.4}^{+0.15}_{-0.17}$ & ${0.4}^{+0.45}_{-0.38}$ & 10.08 \\[3pt]
J1024$-$0719 & DM &  X & X & X & 15 & ${-11.79}^{+0.31}_{-0.41}$ & ${2.47}^{+1.68}_{-1.28}$ & 10.08 \\[3pt]
J1455$-$3330 & RN &  80 & ${-13.28}^{+0.27}_{-1.74}$ & ${1.52}^{+4.59}_{-1.37}$ & X & X & X & 9.27 \\[3pt]
J1600$-$3053 & \textbf{DM+SV} & X & X & X & 97 & ${-12.32}^{+0.5}_{-0.78}$ & ${4.51}^{+2.15}_{-1.5}$ & 9.91 \\[3pt]
J1640+2224 & DM &  X & X & X & 47 & ${-11.57}^{+0.17}_{-0.21}$ & ${0.38}^{+0.75}_{-0.36}$ & 10.22 \\[3pt]
J1713+0747 & RN+DM & 10 & ${-13.73}^{+0.37}_{-0.62}$ & ${1.79}^{+2.25}_{-1.42}$ & 73 & ${-12.17}^{+0.13}_{-0.13}$ & ${1.36}^{+0.51}_{-0.44}$ & 10.08 \\[3pt]
J1730$-$2304 & DM &  X & X & X & 11 & ${-11.91}^{+0.49}_{-0.75}$ & ${2.57}^{+2.6}_{-2.04}$ & 9.93 \\[3pt]
J1738+0333 & \textbf{DM} &  X & X & X & 24 & ${-11.2}^{+0.26}_{-0.57}$ & ${1.84}^{+2.09}_{-1.09}$ & 9.98 \\[3pt]
J1744$-$1134 & \textbf{DM} &  X & X & X & 87 & ${-11.71}^{+0.12}_{-0.11}$ & ${1.13}^{+0.46}_{-0.36}$ & 9.75 \\[3pt]
J1751$-$2857 & DM &  X & X & X & 25 & ${-11.0}^{+0.21}_{-0.61}$ & ${1.06}^{+2.67}_{-0.99}$ & 9.43 \\[3pt]
J1801$-$1417 & DM &  X & X & X & 11 & ${-10.79}^{+0.35}_{-0.37}$ & ${1.64}^{+1.65}_{-1.3}$ & 9.74 \\[3pt]
J1804$-$2717 & DM &  X & X & X & 23 & ${-11.15}^{+0.2}_{-1.66}$ & ${0.58}^{+4.68}_{-0.55}$ & 9.54 \\[3pt]
J1843$-$1113 & DM &  X & X & X & 95 & ${-11.06}^{+0.1}_{-0.1}$ & ${2.37}^{+0.55}_{-0.43}$ & 10.08 \\[3pt]
J1857+0943 & DM &  X & X & X & 20 & ${-12.55}^{+0.62}_{-0.66}$ & ${4.92}^{+1.87}_{-1.92}$ & 9.98 \\[3pt]
J1909$-$3744 & RN+DM & 14 & ${-15.13}^{+0.86}_{-0.64}$ & ${5.72}^{+1.21}_{-2.2}$ & 95 & ${-11.89}^{+0.12}_{-0.1}$ & ${1.63}^{+0.36}_{-0.31}$ & 9.04 \\[3pt]
J1910+1256 & DM &  X & X & X & 20 & ${-12.11}^{+0.75}_{-0.9}$ & ${4.47}^{+2.31}_{-2.32}$ & 9.91 \\[3pt]
J1911+1347 & DM &  X & X & X & 15 & ${-12.14}^{+0.49}_{-1.31}$ & ${3.5}^{+2.71}_{-1.46}$ & 9.92 \\[3pt]
J1918$-$0642 & DM &  X & X & X & 75 & ${-12.34}^{+0.74}_{-1.04}$ & ${4.24}^{+2.44}_{-2.16}$ & 10.08 \\[3pt]
J2124$-$3358 & DM &  X & X & X & 24 & ${-11.47}^{+0.22}_{-1.75}$ & ${1.23}^{+4.17}_{-1.01}$ & 9.68 \\[3pt]
J2322+2057 & NONE & X & X & X & X & X & X & 9.68 \\[3pt]
\bottomrule
\end{tabular*}%
\end{table*}%

For the DR2full+ data, we found that RN, DM and RN+DM are favoured for three, thirteen and eight pulsars respectively, and no significant time-correlated noise was evident in PSR J2322+2057. The data for fourteen pulsars do not support the inclusion of achromatic red noise, and this is further investigated in Section \ref{Perf}. However, including these pulsars in a GWB search analysis is likely still important for the optimisation of the PTA sensitivity as we include the spatial correlations among pulsars. 
For these pulsars, estimates of DM variations in a multi-pulsar analysis with spatial correlations are largely consistent with single-pulsar results with a few exceptions.

Of the 21 pulsars with significant DM variations, twelve have constrained power laws with a spectral index consistent with the expected $\gamma_{\mathrm{DM}} = 8/3$ from Kolmogorov turbulence in IISM. Of the remaining nine pulsars, seven of them also include an achromatic red noise component in the favoured model, which could potentially impact the measurement of DM variations. 

\begin{figure*}
    \centering
    \includegraphics[width=1\columnwidth]{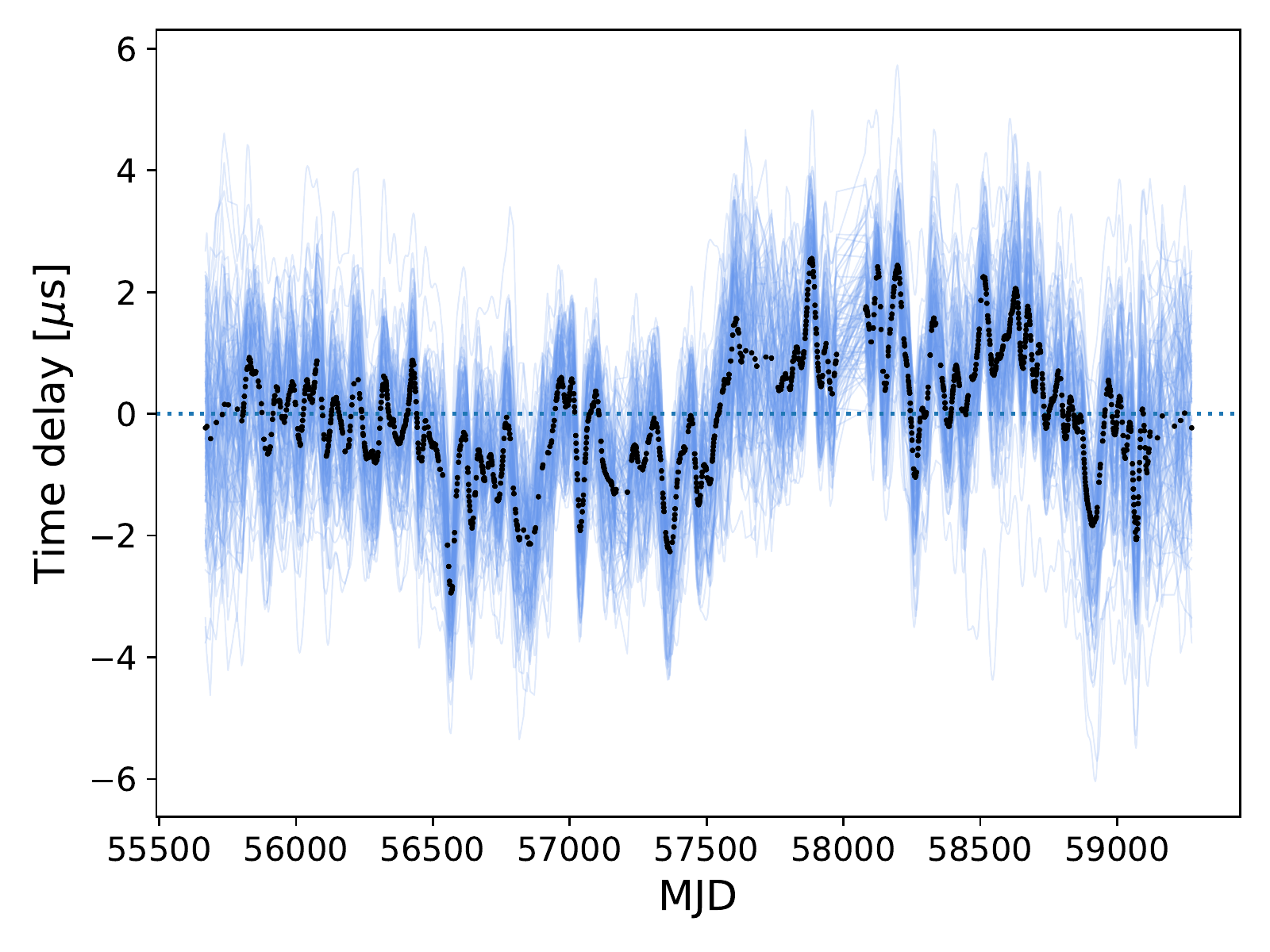}
    \includegraphics[width=1\columnwidth]{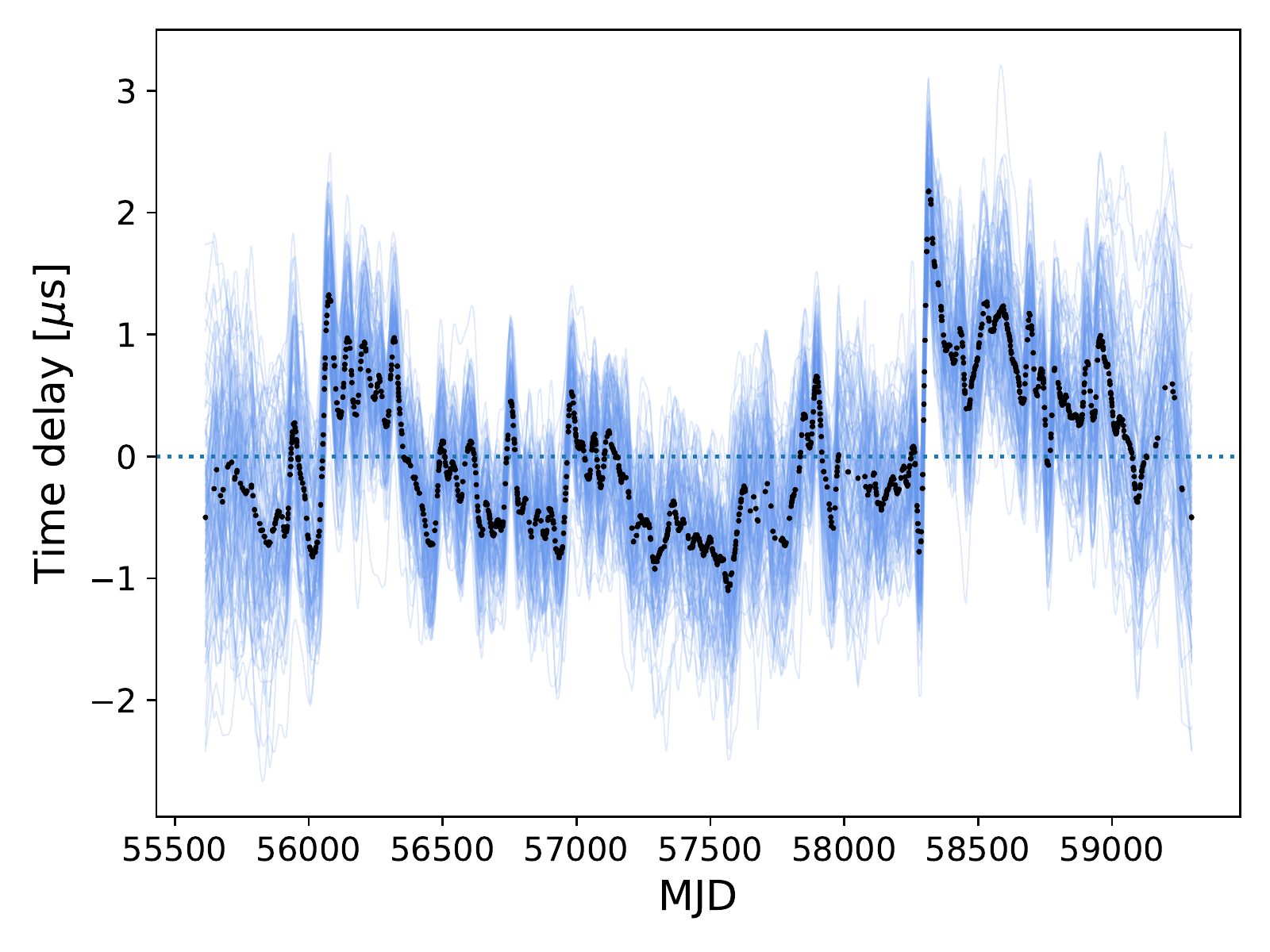}
    \caption{$100$ random realisations (blue) and medians for each epoch (black) of the RN time-delay reconstructions for PSRs J0900$-$3144 (left) and J1012+5307 (right) using the most favoured noise models with the DR2new dataset. The short-timescale stochastic signals seen at a $\mu$s level for these two pulsars still have unknown origins.}
    \label{ShortGammaTimeSerie}
\end{figure*}

We report a significant measurement of achromatic red noise for eleven pulsars, with six consistent with the predicted spectral index $\gamma_{\mathrm{GWB}} = 13/3$ from an idealistic GWB produced from GW-driven circular supermassive black hole binaries (SMBHBs). We find peculiar achromatic red noise for PSRs J0900$-$3144 and J1012+5307, with flat power-laws (resp. $\gamma_{\mathrm{RN}} = 1.06^{+0.28}_{-0.27}$ and $\gamma_{\mathrm{RN}} = 1.21^{+0.17}_{-0.17}$), displaying short-term correlated noise, as shown in Figure \ref{ShortGammaTimeSerie}.

For DR2new, we find significant achromatic noise only for seven pulsars, four of which are consistent with $\gamma_{\mathrm{GWB}}=13/3$. The lower number of pulsars favouring RN as compared to DR2full+ can either be due to the shorter timing baselines which reduce the sensitivity for long-term red noise signals or due to unmodeled noise present in the first half of the dataset. Of the 22 pulsars with significant DM, sixteen are consistent with $\gamma_{\mathrm{DM}}=8/3$. Of the remaining six,  three pulsars have RN+DM as the favoured model. PSRs J1640+2224 and J1744$-$1134 display unusually low spectral indices at $0.38^{+0.75}_{-0.36}$ and $1.13^{+0.46}_{-0.36}$ respectively. Conversely, we report high DM variations spectral index for PSR J1600$-$3053 at $4.51^{+2.15}_{-1.5}$, for which the favoured noise model is DM+SV. More details on the noise properties for this pulsar are described in Section \ref{InPTA}. Figures for posterior distributions, time-domain realisations, and relevant tables generated for each pulsar are available on Zenodo \footnote{10.5281/zenodo.802501}.

\section{Interpretation}
\label{interpretation}

\subsection{Comparison of EPTA Noise Models}
\label{noisecompare}
In this section, we compare the noise models between DR2full, DR2new and the original models reported for the EPTA DR1 dataset from \citet{cll+16}. We find that the pulsars largely fall into three categories:
\begin{itemize}
    \item \textbf{Category1: Consistent noise models in the three datasets}: Typically, DR2full improves over DR1 or DR2new, and both DR2full and DR2new do a better job of constraining chromatic noise than DR1.
    \item \textbf{Category2: DR2full and DR2new find chromatic noise}: Several pulsars in DR1 found only achromatic noise or were unable to distinguish between achromatic and DM noise, but the DR2full and DR2new datasets prefer models with only DM variations. 
    \item \textbf{Category3: Complex noise models}: In a handful of pulsars, including several of those with the best TOA precision, we find a more complex relationship between the noise models in the three datasets.
\end{itemize}

\begin{figure*}
    \centering
    \begin{tabular}{cc}
        \includegraphics[width=1\columnwidth]{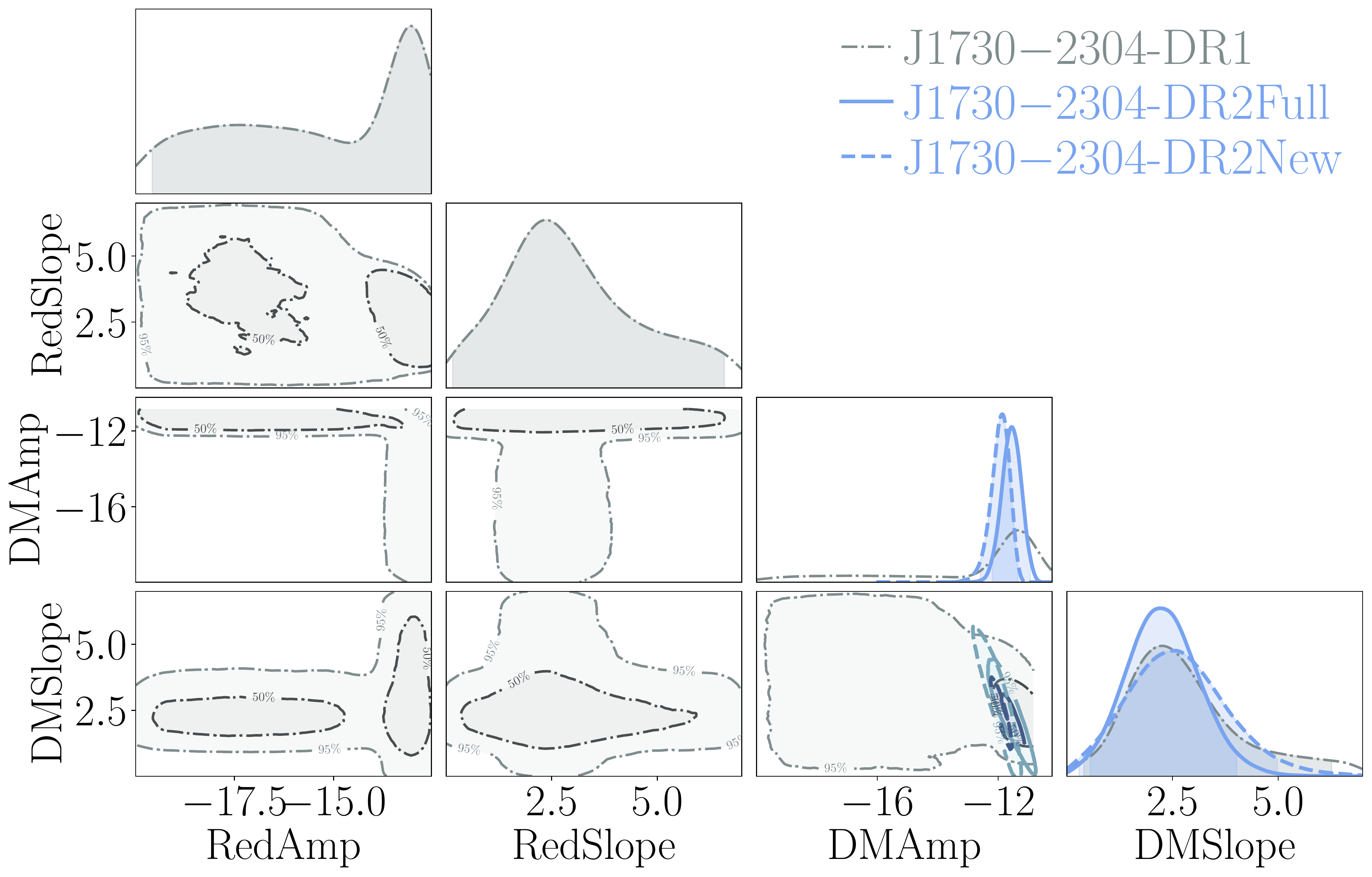}     & \includegraphics[width=1\columnwidth]{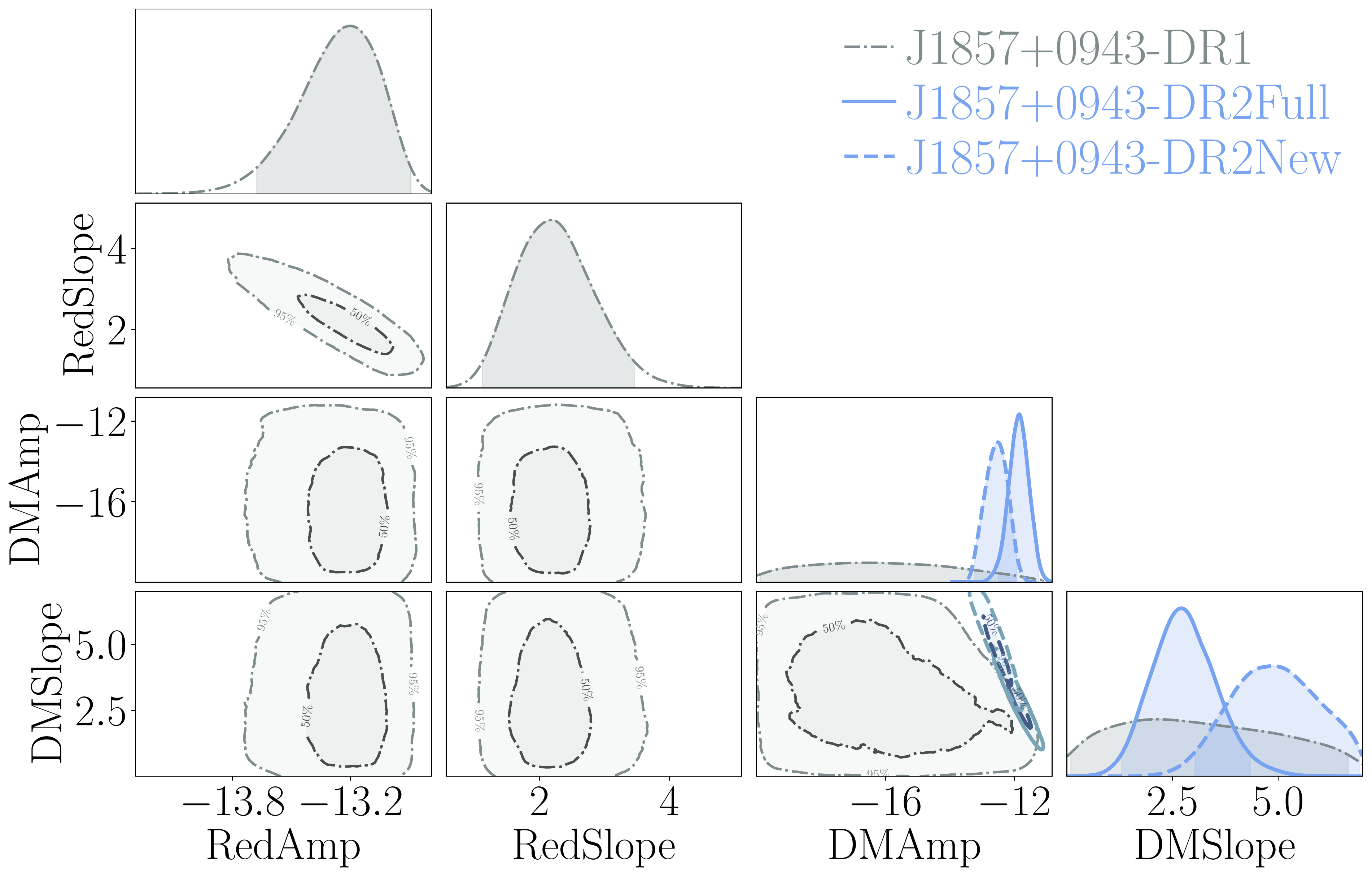} \\
        \includegraphics[width=1\columnwidth]{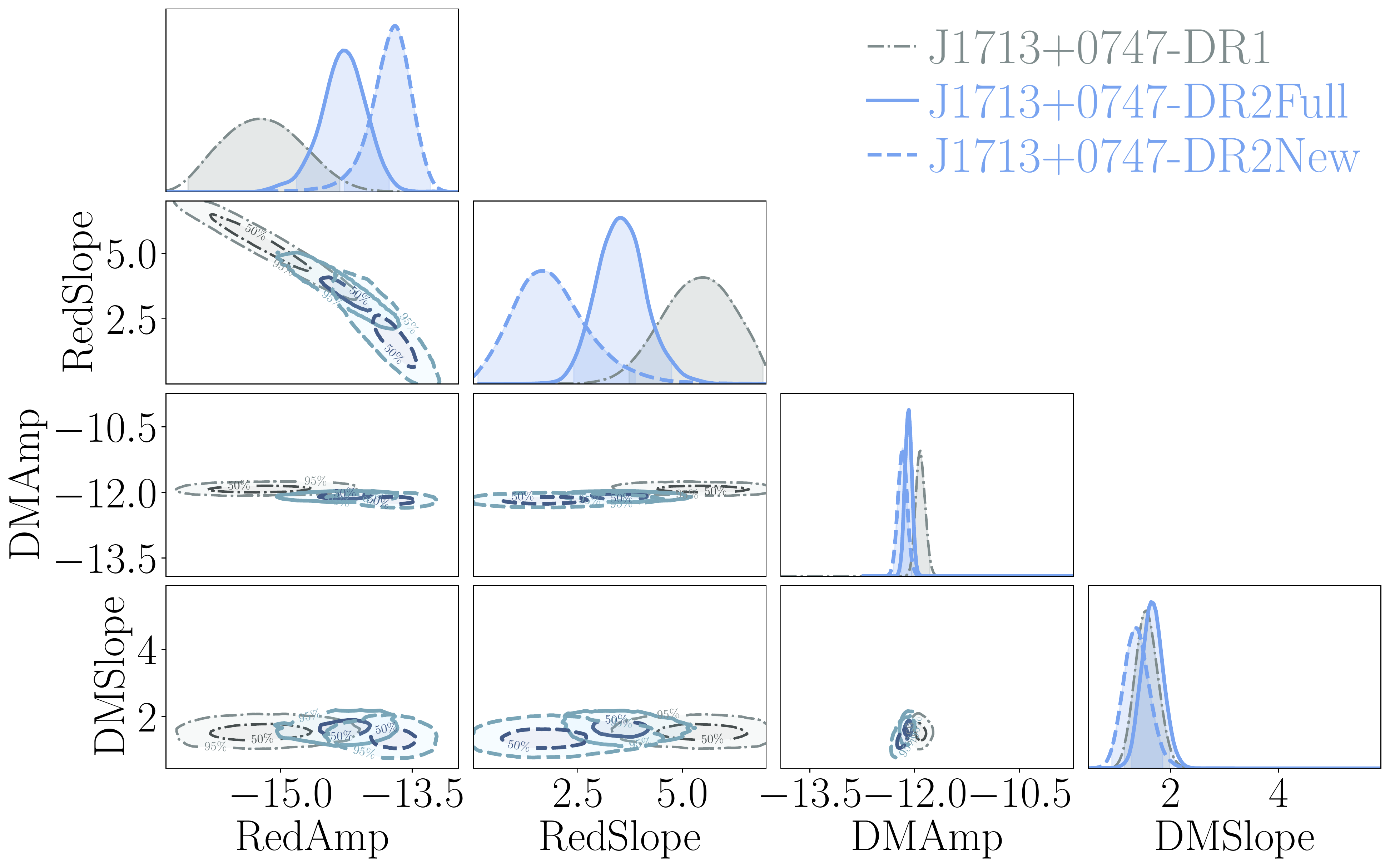}     &   \includegraphics[width=1\columnwidth]{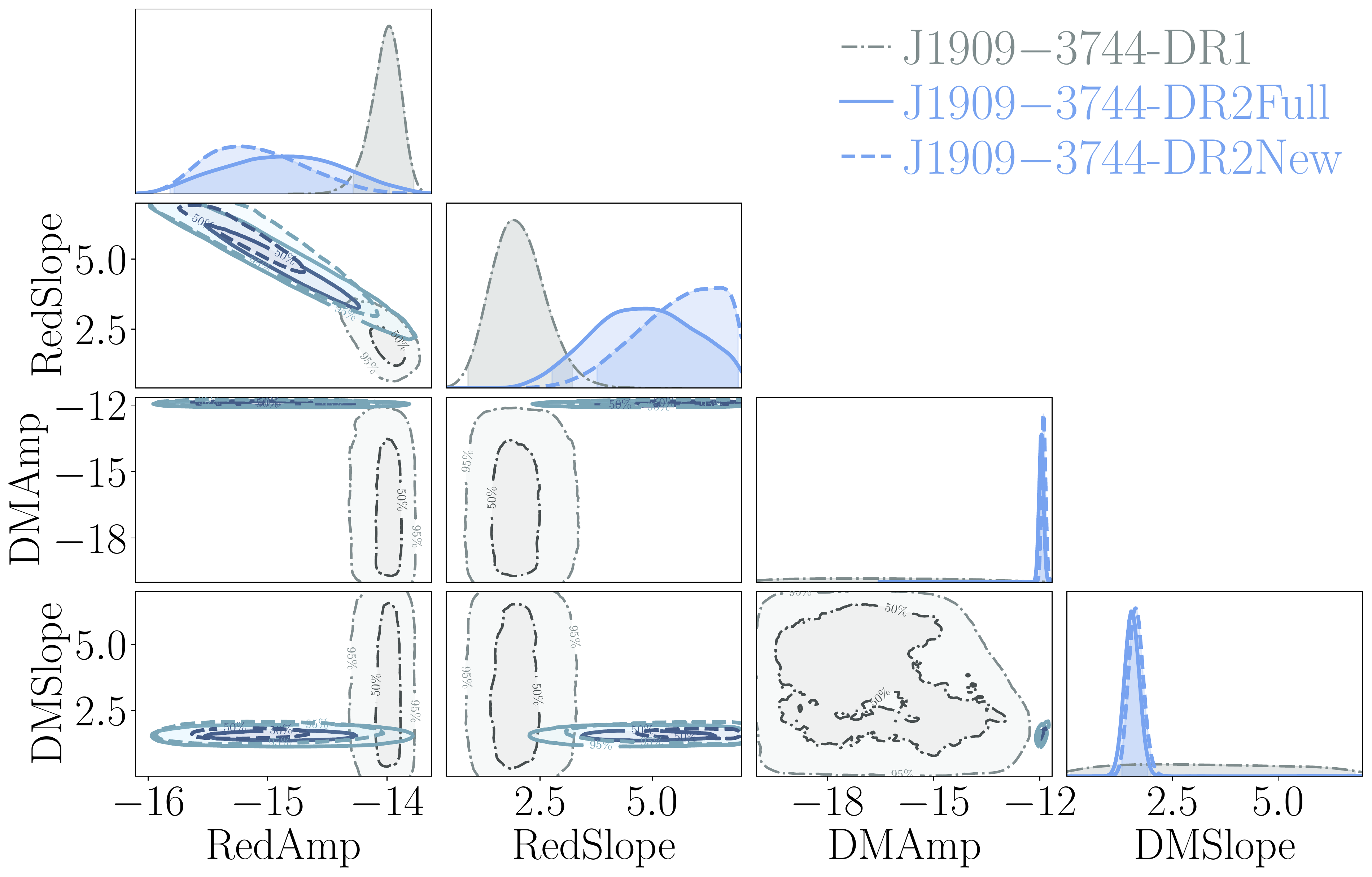} \\
    \end{tabular}
    \caption{(From top left to bottom right) (a) PSR J1730$-$2304 falls in Category 1, where the DR2full/New datasets are able to constrain chromatic noise (DMAmp, DMSlope). (b) For PSR J1857+0943, belonging to Category 2, it is clear that the DR1 data could not disentangle chromatic from achromatic noise (RedAmp, RedSlope), which is resolved better with the DR2 datasets. (c) PSR J1713+0747 in Category 3, shows clear inconsistencies in the achromatic noise estimates and (d) PSR J1909$-$3744 shows a much more complicated behaviour as described in the text motivating its inclusion in Category 3.}
    \label{dr1dr2noisecompare}
\end{figure*}

In Figure \ref{dr1dr2noisecompare}, we present noise models for four pulsars as an example of belonging to either of the above categories. Below, we present a summary of the noise models in the three datasets for each of the 25 pulsars. 

\textbf{PSR J0030+0541 (Category 1)} -- DR2full and DR2new largely agree, both with achromatic noise only. The DR2 posterior is better constrained. DR1 finds the same achromatic noise, but is unable to cleanly distinguish it from the DM noise.

\textbf{PSR J0613$-$0200 (Category 3)} -- DR2 and DR1 both see achromatic noise of similar amplitude, but the spectral index is unconstrained for DR1, whereas DR2 finds the achromatic noise to have an index $>4$. DR2new is unable to detect this noise, probably due to the relatively short data ($\sim$ 10 years), and only finds DM variations. The posterior for $\gamma_\mathrm{DM}$ shifts somewhat between the three datasets but are largely consistent and slightly flatter than Kolmogorov.

\textbf{PSR J0751+1807 (Category 1)} -- All three datasets are largely consistent, finding only DM variations with a spectral slope consistent with Kolmogorov.

\textbf{PSR J0900$-$3144 (Category 3)} -- DR2 and DR2new posteriors are consistent, finding achromatic noise with a very flat spectral shape $<2$, and DM variations with an unconstrained spectral slope. The achromatic noise is much flatter than expected from classical power-law timing noise and may represent another noise process. The chromatic noise in DR1 data is consistent but unable to discern the achromatic noise.

\textbf{PSR J1012+5307 (Category 3)} -- DR2full has a tightly constrained posterior for DM and achromatic noise. The achromatic index is very flat $<2$ and appears to be dominated by the high-frequency power above $1/\mathrm{yr}$. DR2new finds similar achromatic noise, and similar amplitude DM variations, but is barely able to constrain $\gamma_\mathrm{DM}$. We speculate that this is because the longer dataset of DR2full can rule out very steep DM variations given the absence of steep achromatic noise. DR1 finds achromatic noise with a marginally steeper spectrum and finds an upper limit on DM variations just below that observed in the DR2full data.

\textbf{PSR J1022+1001 (Category 2)} -- DR2full and DR2new are consistent with each other and show a very flat spectrum DM variation and an achromatic noise with $\gamma_\mathrm{red} \sim 4$. The flat-spectrum DM variation in this pulsar could likely be due to high-frequency residuals remaining from the solar wind contribution as this pulsar passes close to the ecliptic. A revised interpretation is given in Section \ref{Perf}, after performing a chromatic index evaluation for each time-correlated component. DR1 showed flat spectrum achromatic noise and little variation in DM. We suspect that this may be a leakage from DM variations due to the limited frequency coverage. We account for the solar wind by fitting for the \texttt{NE-SW} parameter as part of the timing model, which results in an estimated value of $10.9\pm0.3 \ \mathrm{cm}^{-3}$ relative to the constant value of $7.9 \ \mathrm{cm}^{-3}$ in the other pulsars. 

\textbf{PSR J1024$-$0719 (Category 2)} -- DR1 showed a marginal detection of DM variations with a steep spectral index. DR2full and DR2new both find only DM variations consistent with Kolmogorov and incompatible with the DR1 value. We find that the DR2full and DR2new models seem more plausible, but they could indicate the presence of smooth DM structures in the early DR1 data.

\textbf{PSR J1455$-$3330 (Category 1)} -- The three datasets are largely consistent. DR1 cannot distinguish between DM and achromatic noise, but the better frequency coverage of DR2full and DR2new finds only achromatic noise. The DR2full posteriors are more tightly constrained, particularly in $\gamma_\mathrm{red}$.

\textbf{PSR J1600$-$3053 (Category 3)} -- DR1 has fairly flat spectrum DM variations, which are split into flat-spectrum scattering delay variations and steep (than Kolmogorov) DM variations in DR2full and DR2new.

\textbf{PSR J1640+2224 (Category 2)} -- DR1 is unable to distinguish between DM and achromatic noise. DR2full and DR2new find only DM variations with an extremely flat spectral index, $\gamma_\mathrm{DM}<1$.  It is not clear what physical process would give rise to such a flat spectral index for DM variations.

\textbf{PSR J1713+0747 (Category 3)} -- Two DM events (exponential dips) at MJDs $\sim 54757$ and one at MJD $\sim 57510$ are observed for this pulsar in DR2full. The methodologies adopted to model these events are different between DR1 and DR2, but we find that this does not affect the red noise properties. However, from Figure \ref{dr1dr2noisecompare}, it is clear that the red noise properties change considerably between DR1, DR2new, and DR2full with the red noise spectral index going from steep to shallow, respectively. We think that this could likely point to either our assumptions of stationarity being incorrect or processes not well-modelled by a power law. 

\textbf{PSR J1730$-$2304 (Category 1)} -- Generally consistent between the three datasets, with DR1 unable to distinguish between DM and achromatic noise. DR2full and DR2new find only DM variations consistent with Kolmogorov.

\textbf{PSR J1738+0333 (Category 2)} -- DR1 does not find any noise. DR2full and DR2new find a flat-spectrum noise process, but the model selection prefers achromatic noise for DR2full and chromatic noise for DR2new. In both cases, the evidence is marginal for any one particular model.

\textbf{PSR J1744$-$1134 (Category 3)} -- The three datasets are superficially similar, but paint a confusing picture when taken together. DR1 finds no variations in DM and an achromatic process with $\gamma_\mathrm{red}\sim 3$. DR2full finds DM variations with $\gamma_\mathrm{DM}<1$, amplitude above the upper limit for DM variations from DR1, and achromatic noise of similar amplitude but with a steeper, but largely unconstrained spectral slope, $\gamma_\mathrm{red} > 2.5$. Indeed, the DR2full posterior for the achromatic noise is bimodal with a component that is largely consistent with DR1 and a component that is much steeper ($\gamma_\mathrm{red} > 5$). DR2new does not find evidence of achromatic noise, but has consistent DM variations with DR2full. The very flat spectrum DM variations, inconsistent with DR1, suggest that the chromatic noise is not a power-law process but instead dominated by high-frequency variations to which DR1 was insensitive because of the limited instantaneous frequency coverage. The inconsistency of the achromatic noise may indicate there is another underlying process at work in this pulsar beyond a single power-law achromatic noise and DM variations. This pulsar is observed by all members of the IPTA, and hence we suspect that the picture will become clearer when studied with a combined IPTA dataset.

\textbf{PSR J1751$-$2857 (Category 1)} -- Generally consistent between the three datasets. DR1 only finds upper limits, but DR2full and DR2new find a DM variation consistent with Kolmogorov. 

\textbf{PSR J1801$-$1417 (Category 1)} -- Generally consistent between the three datasets. DR2full and DR2new better constrain the DM variation, and the spectral shape is consistent with Kolmogorov.

\textbf{PSR J1804$-$2717 (Category 1)} -- Generally consistent between the three datasets. DR2full and DR2new much better constrain the DM variations, and the spectral shape is consistent with Kolmogorov, although both prefer a flatter $\gamma_\mathrm{DM}$.

\textbf{PSR J1843$-$1113 (Category 1)} -- Generally consistent between the three datasets. DR2full and DR2new much better constrain the DM variation, and the spectral shape is consistent with Kolmogorov.

\textbf{PSR J1857+0943 (Category 2)} -- DR1 finds achromatic noise that DR2full and DR2new attribute to DM variations, without any achromatic noise. There is some overlap in the DM posteriors for DR1 and DR2full, though we suspect that the detection of achromatic noise in DR1 was incorrect.

\textbf{PSR J1909$-$3744 (Category 3)} -- DR1 finds achromatic noise that seems to have the same properties as the DM variations in the DR2full/DR2new data. The DM variations in DR2full/DR2new are above the upper limit from DR1. DR1 had minimal frequency coverage for this pulsar, though it is not clear why achromatic noise was preferred if there was a degeneracy. DR2full and DR2new find achromatic noise with a steeper spectral index and at a lower amplitude than the DM variations in the DR1 data. Note that the noise analysis in \citet{lgi+20} which used very similar EPTA data as in this work, reported similar results when the L-band data were divided into sub-bands.

\textbf{PSR J1910+1256 (Category 1)} -- Generally consistent between the three datasets. DR1 only finds upper limits, while DR2full and DR2new find DM variations consistent with Kolmogorov, but $\gamma_\mathrm{DM}$ is not well constrained.

\textbf{PSR J1911+1347 (Category 2)} -- Generally consistent between the three datasets. DR1 is not able to distinguish between DM and achromatic noise, but DR2full and DR2new find a DM variation consistent with Kolmogorov.

\textbf{PSR J1918$-$0642 (Category 2)} -- DR1 finds achromatic noise with a preference for large $\gamma_\mathrm{red}$, and does not find significant DM variations. DR2full does not find evidence for achromatic noise, but instead finds a DM variation consistent with Kolmogorov, although also consistent with a steeper spectrum. DR2new is consistent with DR2full. This is another example of achromatic noise in DR1 being interpreted as chromatic noise when wide-band receivers are used.

\textbf{PSR J2124$-$3358 (Category 2)} -- Generally consistent between the three datasets. DR1 only finds upper limits, but DR2full and DR2Bew find a variation in DM consistent with Kolmogorov, though preferring a flatter $\gamma_\mathrm{DM}$.

\textbf{PSR J2322+2057} -- Does not show evidence of time-correlated noise in any EPTA dataset.

\subsection{Changes in noise models after the inclusion of the InPTA data}
\label{InPTA}
Here we study the impact of including low radio frequency observations from the InPTA on the estimated noise models. The InPTA dataset complements the EPTA data with simultaneous observations at 300-500 MHz and at 1260-1460 MHz between MJDs 58235 and 59496 observed with the upgraded Giant Metrewave Radio Telescope (uGMRT). The frequency coverage at 300-500 MHz is particularly important since EPTA has a limited number of observations at this frequency band. Therefore, the inclusion of the InPTA data is of particular interest in constraining noise due to the IISM such as DM and scattering variations. To allow a quantitative comparison of the posterior noise models before and after the inclusion of InPTA data, we adapt and employ a tension metric package as detailed in \cite{raveri2021non}\footnote{https://github.com/mraveri/tensiometer}. 

We begin by introducing the probability density for parameter differences between the posteriors of two sets of parameters $\boldsymbol \theta_1$ and $\boldsymbol \theta_2$ to be $\mathcal{P}(\Delta \boldsymbol \theta)$ where $\Delta \boldsymbol \theta \equiv \boldsymbol \theta_1-\boldsymbol \theta_2$. Employing a joint posterior $\mathcal{P}(\boldsymbol \theta_1,\boldsymbol \theta_2) \equiv \mathcal{P}(\boldsymbol \theta_1,\boldsymbol \theta_2|d_1,d_2) $,  we can write $\mathcal{P}(\boldsymbol \theta_1,\Delta \boldsymbol \theta)=\mathcal{P}(\boldsymbol \theta_1,\boldsymbol \theta_1-\Delta \boldsymbol \theta)$, where $d_1$ and $d_2$ refer to the two data sets.  Integrating the base parameters, we obtain

\begin{align}
\label{eqdef}
    \mathcal{P}(\Delta \boldsymbol \theta)=\int_{V_\pi} \mathcal{P}(\boldsymbol \theta_1,\boldsymbol \theta_1 -\Delta \boldsymbol \theta) \, d \boldsymbol \theta_1\,,
\end{align}

where $ V_\pi $ is the volume of the parameter space where the prior is non-vanishing, also known as the support of the prior. If the two datasets, specified by the parameter $\boldsymbol \theta$, are conditionally independent,  it is possible to sample the posteriors $\mathcal{P}(\theta_1)$ and $\mathcal{P}(\theta_2)$ separately \citep{raveri2021non}. In this scenario, the expression for 
$\mathcal{P}(\Delta \boldsymbol \theta)$ becomes 
\begin{align}
    \mathcal{P}(\Delta \boldsymbol \theta)=\int_{V_\pi} \mathcal{P}(\boldsymbol \theta_1) \, \mathcal{P}(\boldsymbol \theta_1 -\Delta \boldsymbol \theta) \, d \boldsymbol \theta_1 \ .
\end{align}

Thereafter, we obtain the mean probability of the presence of a parameter shift using the following equation. 
\begin{align} 
\label{eqten}
    \Delta=\int_{\mathcal{P}(\Delta \boldsymbol \theta) > \mathcal{P}(0)} \mathcal{P}(\Delta \boldsymbol \theta) d \Delta \boldsymbol \theta, 
\end{align}

which incorporates the posterior mass above the iso-contour of no shift. We then convert the above $\Delta$ into an effective number of $\sigma$ using the standard normal distribution. Detailed comparisons that arise from the posteriors of DR2full and DR2full+ are available in the following URL\footnote{https://github.com/subhajitphy/Posterior\_comparisons}. In Table \ref{EPTA_InPTA_tension}, we report the estimated tension
(in $\sigma$) for the red and DM noise models while dealing with
DR2full and DR2full+ datasets. It shows that the 2D posterior distributions of the RN and DM parameters are consistent ($\Delta < 1\sigma$) for all parameters, except for the power law DM variations of the PSRs J0613$-$0200, J1600$-$3053, J1744$-$1134 and J1909$-$3744. In the following, we discuss possible explanations for these pulsars. 

\begin{table}
    \caption{Estimated tension (Z-score in sigma) between the DR2full and DR2full+ datasets for the red and DM noise models. Instances with significant tension are highlighted.}
    \centering
    \begin{tabular}{l|l|l|l}
    \hline
        Pulsar & Model & RN-RN & DM-DM \\ \hline
        J0613$-$0200 & DM+RN & 0.74 & \textbf{2.97} \\
        J0751+1807 & DM & X & 0.63 \\ 
        J1012+5307 & DM+RN & 0.02 & 0.04 \\ 
        J1022+1001 & DM+RN & 0.08 & 0.52 \\ 
        J1600$-$3053 & DM & X & \textbf{4.64} \\ 
        J1713+0747 & DM+RN & 0.01 & 0.14 \\ 
        J1744$-$1134 & DM+RN & 0.20 & \textbf{2.29} \\
        J1857+0943 & DM & X & 0.05 \\ 
        J1909$-$3744 & DM+RN & 0.05 & \textbf{4.39} \\
        J2124$-$3358 & DM & X & 0.84 \\ \hline
    \end{tabular}
    \label{EPTA_InPTA_tension}
\end{table}

\subsubsection{PSR J0613$-$0200}
\begin{figure*}
    \centering
    \includegraphics[width=1\columnwidth]{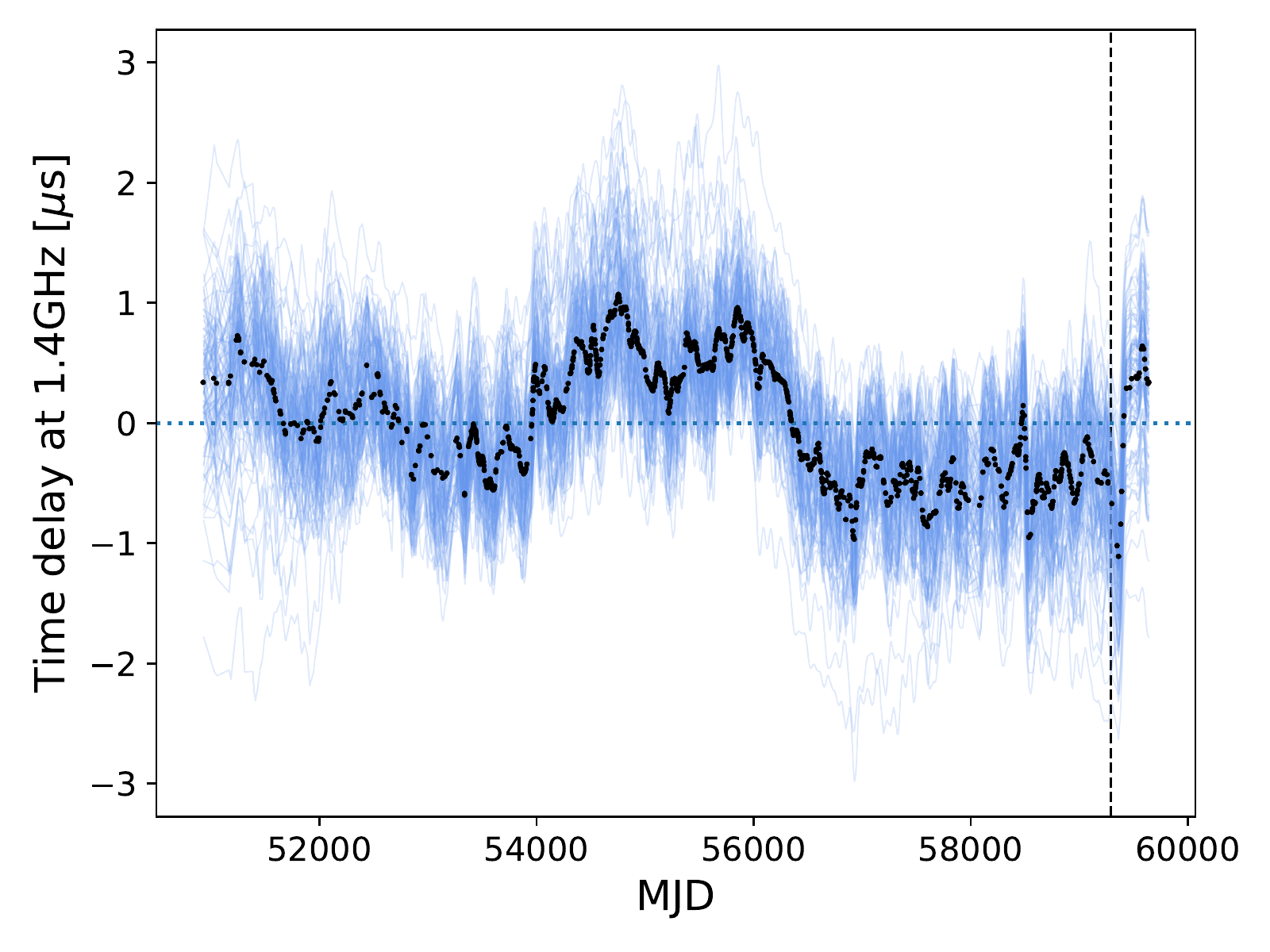}
    \includegraphics[width=1\columnwidth]{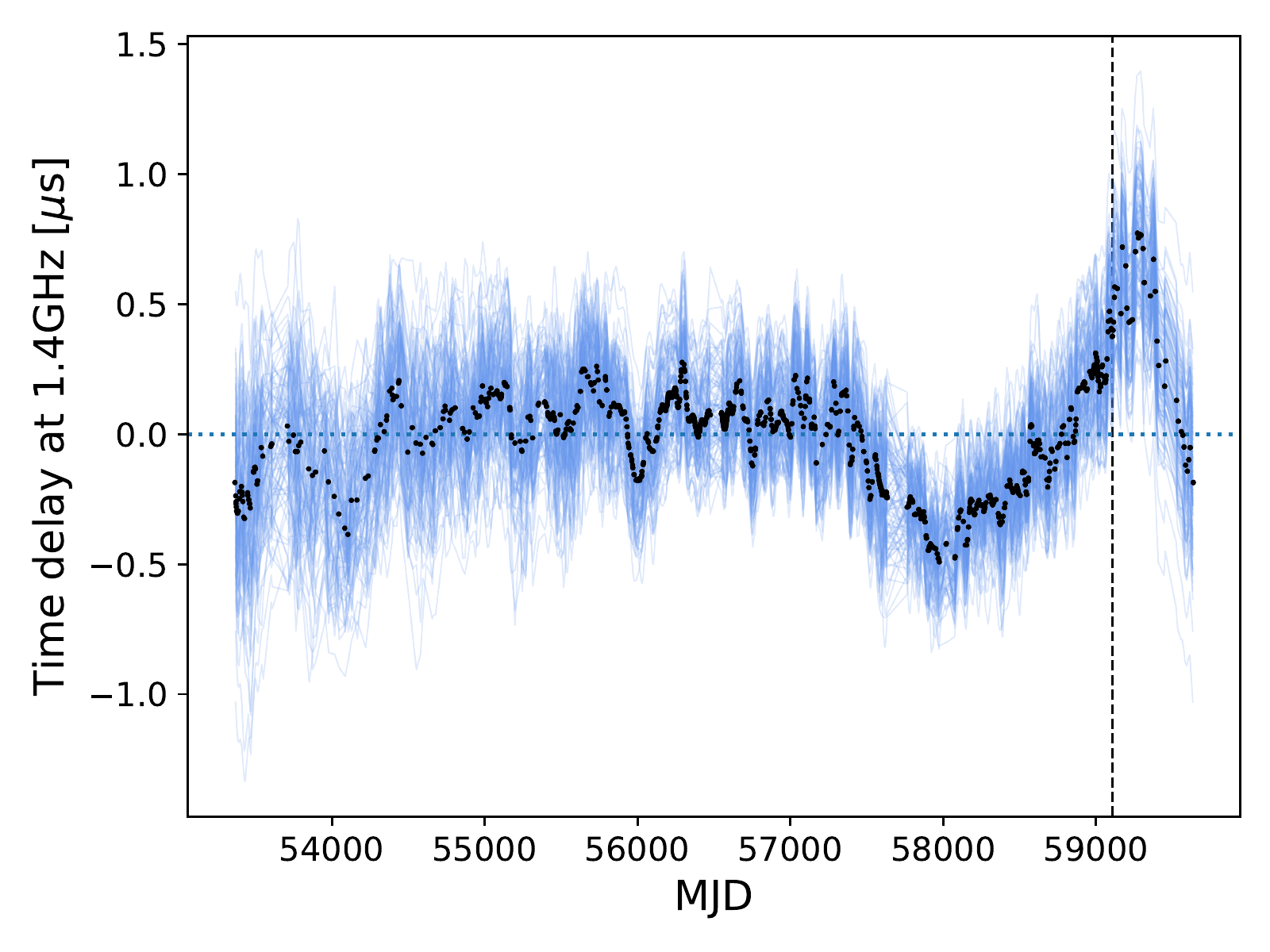}
    \caption{$100$ random realisations (blue) and medians for each epoch of the time-domain reconstructions of the DM variations delays at $1.4$ GHz radio frequency modelled as a Gaussian process for PSRs J0613$-$0200 (left) and J1909$-$3744 (right), with the ``DR2full+''. The black dashed lines display the last epochs of DR2full (i.e. EPTA only). The inclusion of InPTA data allows to measure sharp changes at after MJD $59000$. The delays are obtained from marginalizing the timing model parameters which comprise the DM constant and the first two time derivatives of the dispersion measure (DM1, DM2).}
    \label{J0613_timeserie}
\end{figure*}

For this pulsar, combining InPTA with the EPTA data yields a lower spectral index and higher amplitude at $f_{\mathrm{yr}}$ for the chromatic noise.
We observe an interesting sharp jump in the last years of the DM time series after including InPTA data (cf. the left panel of Figure \ref{J0613_timeserie}), which might increase the power at high PSD frequencies, and therefore yield to a flatter constrained power law.

\subsubsection{PSR J1600$-$3053}
\begin{figure*}
    \centering
    \includegraphics[scale=0.44]{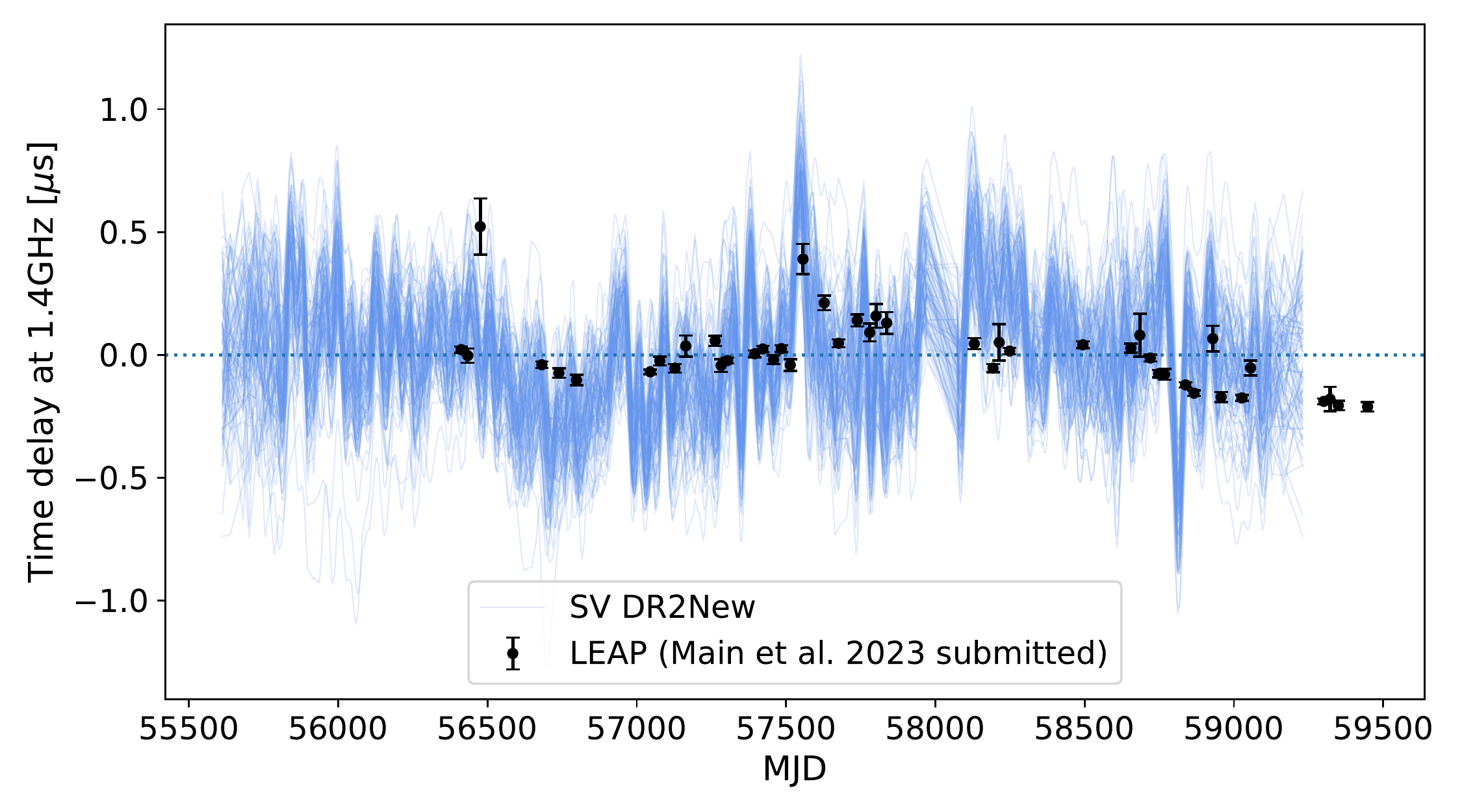}
    \includegraphics[scale=0.36]{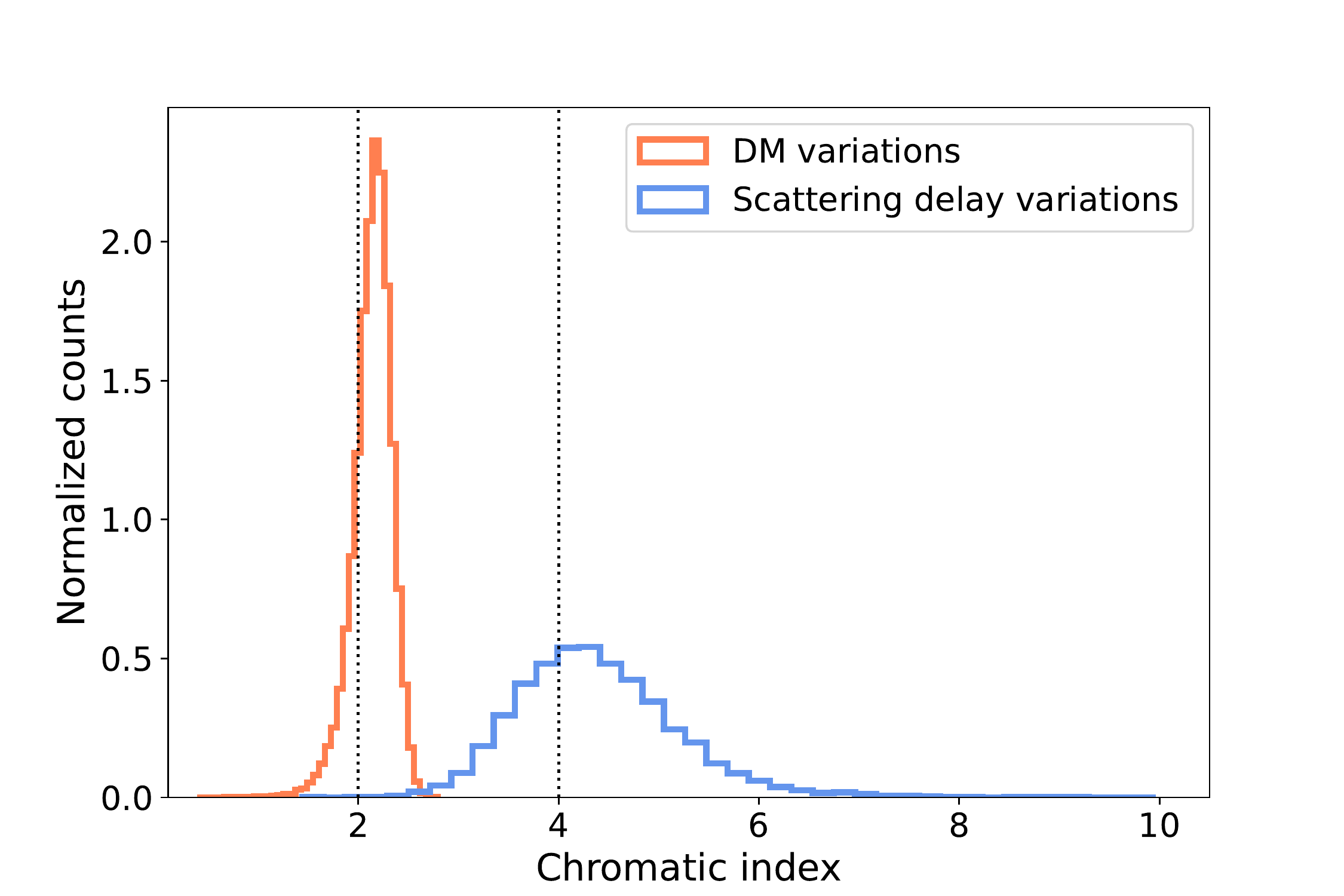}
    \caption{Scattering delay variations measurement for PSR J1600$-$3053 with DR2new. Left plot: $100$ random realisations (blue) of the time-domain reconstructions of the scattering delay variations at $1.4$ GHz radio frequency modelled as a Gaussian process with "DR2new". Independent scattering delays measured from scintillation analysis with the LEAP data \citep{mac+23} are shown by the black points. Right plot: Chromatic index posterior distributions measured for the DM (orange) and the SV (blue) with "DR2new", while fixing the chromatic index of the other component. The favoured model for this dataset includes both DM and SV. The black dotted lines emphasise the expected values for DM and SV, respectively at $2$ and $4$.}
    \label{J1600_timeserie}
\end{figure*}

The favoured model using DR2full comprises DM and SV, with constraints that are (1) consistent with independent measurements of scattering delays with the Large European Array for Pulsars (LEAP) \citep{mac+23} (cf. left panel of Figure \ref{J1600_timeserie}), and (2) highly consistent with the expected chromatic indices for both DM and SV as shown in the right panel of Figure \ref{J1600_timeserie}. However, the inclusion of InPTA data no longer supports scattering variations, and the favoured model is RN+DM. While the inclusion of SV is still supported after including the L-band data from the InPTA, we find that this discrepancy is happening from including the P-band data ($300 - 500$ MHz). Furthermore, the scattering variations in DR2full are unlikely to be related to pulse broadening at low frequencies, as we have no evidence for this in the InPTA P-band data. This discrepancy will be further investigated with data from the IPTA. 

\subsubsection{PSR J1744$-$1134}
\begin{figure}
    \raggedright
    \includegraphics[width=1\columnwidth]{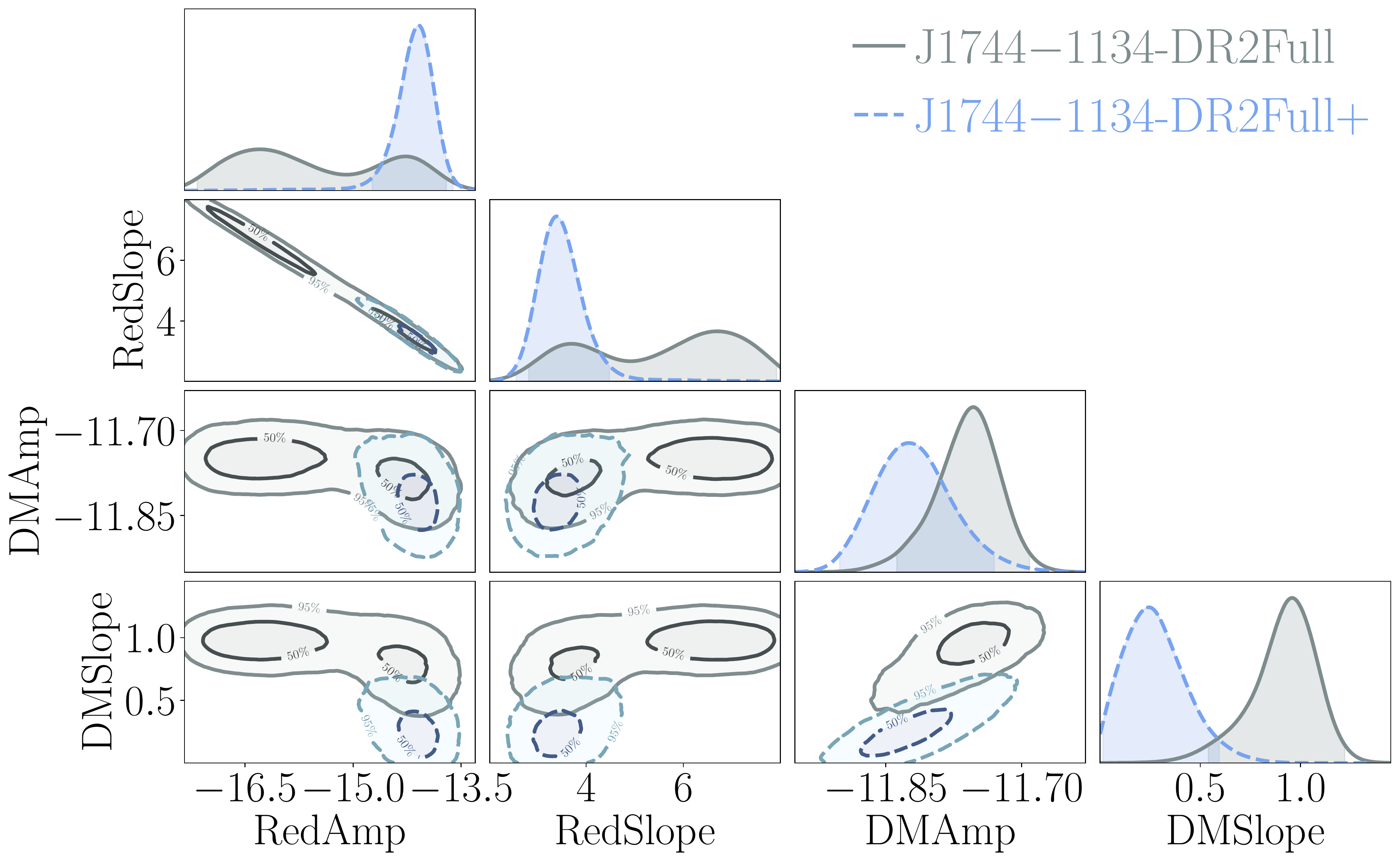}  
    \caption{Red and DM noise models for PSR J1744$-$1134 using DR2full and DR2full+ datasets. The inclusion of InPTA data allows a better constraint on the achromatic noise.}
    \label{1744_inpta}
\end{figure}

As described in Section \ref{noisecompare}, this pulsar exhibits a bimodal posterior for achromatic noise and a very flat spectrum for DM variations with the DR2full dataset. The inclusion of the InPTA data measures DM variations with a much lower spectral index and a lower amplitude. This removes the observed bimodality in the achromatic noise and allows a tighter constraint on a single mode as shown in Figure \ref{1744_inpta}).

\subsubsection{PSR J1909$-$3744}
The behaviour observed in PSR J1909$-$3744 appears to be very similar to PSR J0613$-$0200, in which including the InPTA data produces a lower spectral index and a higher amplitude for the chromatic noise. This is also likely caused by the sharp change in DM variations in the last two years of the dataset (cf. Figure \ref{J0613_timeserie}). This feature was already reported in \citet{tnr+22} and is also shown in \citet{cpb2023} that used, respectively, the first data release of InPTA and the ultra-wide-bandwidth low-frequency data from PPTA. Interestingly, we do not observe any associated impact on the RN posterior distributions. 

\subsection{Implications}
As expected the much-improved frequency coverage of DR2 has meant that the chromatic noise is much better constrained with the DR2full and DR2new datasets. Surprisingly, in some cases, signals attributed to achromatic processes in DR1 have been attributed to chromatic noise in DR2.

Of particular interest, however, are pulsars like J1713+0747 and J1744$-$1134, in which the choice of dataset seems to affect the noise model hyperparameters in a way that is not easily understood. Although these effects are rather subtle, these are two of the best-timed pulsars in the EPTA, and this inconsistency may be an indication that more complex noise models may be needed for the best-timed pulsars and longest datasets. We assume that the noise process is a stationary Gaussian process with a power-law spectral density, and this may suggest that one or more of these assumptions are incorrect. Indeed, it is quite plausible that the pulsar spin noise is not entirely stationary.
The IISM is known to have discrete structures that may change the statistics of the DM and scattering variations. In particular, PSR J1713+0747 is observed to undergo `chromatic timing events that are neither a stationary process nor fully modelled by achromatic or $\nu^{-2}$ DM variations \citep{leg+18}.
Pulsar timing noise in the population of canonical pulsars also shows behaviour that is either not strictly modelled by a power-law \citep[e.g. quasi-periodic variability;][]{lhk+10} or shows discrete changes in spin-down behaviour on long timescales (e.g. \citealp{bkb+14,ssw+22}).

It is worth reiterating that the noise models presented in this work represent our best estimate of the underlying noise using the understanding that we currently have. Further insights are hampered by the fact that it can be hard to disentangle the effect of improvements in the observing systems, especially sensitivity and observing frequency coverage, from changes in the observable pulsar noise properties, either due to time-variable processes or from processes only detectable on long timescales. Combined IPTA datasets bring additional complications, but the combination of improved instantaneous sensitivity and frequency coverage, particularly in the latest instrumentation, will be our best chance to separate instrumental effects from those intrinsic to the pulsars. Furthermore, the direct combination of data can make use of a dropout analysis to identify any telescope- or backend-specific instrumental effects that may mask the real behaviour of the pulsars.

\section{Noise Model Validation}
\label{noisevalidation}

In this section, we perform several tests to compare the performance of the customised noise models with more standard models and assess the robustness of the results. All the related materials with Figures and values can be found in the GitLab repository (Link available soon).

\subsection{Performance of the Customised Noise Models}
\label{Perf}

As shown by Tables \ref{eptatable1} and \ref{eptatable2}, the favoured noise models include only one component for most pulsars, making them simpler than the standard models commonly used in PTA analyses (RN and DM with $30$ and $100$ frequency bins, respectively, for all pulsars). We first evaluate the improvements enabled by the model selection process by comparing them with standard noise models, except for PSR J1012+1001, since we use the standard model for this pulsar (cf. Section \ref{results}). The Bayes factors in favour of the customised noise models over the standard models are shown in columns $3$ of Table \ref{eptatable3} for each dataset. 
Following the \citet{kass_rafferty} scale, we find 
\begin{itemize}
    \item weak support ($\mathcal{B}^{\mathrm{cus}}_{\mathrm{stand}} \in [1, 3]$) for $11$ and $13$ pulsars, 
    \item positive support ($\mathcal{B}^{\mathrm{cus}}_{\mathrm{stand}} \in [3, 20]$) for $5$ and $7$ pulsars,
    \item strong support ($\mathcal{B}^{\mathrm{cus}}_{\mathrm{stand}} \in [20, 100]$) for $2$ and $0$ pulsars,
    \item very strong support ($\mathcal{B}^{\mathrm{cus}}_{\mathrm{stand}} \geq 100$) for $6$ and $4$ pulsars,
\end{itemize}
in DR2full+ and DR2new respectively. 
We notice higher Bayes factor values for DR2full+ compared to DR2new for most pulsars, which can be partially explained by the better performance of the standard noise models to describe higher PSD frequencies for DR2new as its timing baseline is $\sim$ two times shorter than DR2full+. Furthermore, cases with ``strong'' and ``very strong'' evidence are obtained for the pulsars that include two time-correlated components (RN+DM, DM+SV), except for PSR J1801$-$1417 (only DM variations). We also observe a very significant reduction of Bayes factors for PSR J1909$-$3744 with the DR2new, even if the preferred model for this pulsar includes both RN and DM for the two datasets.

To assess the validity of the customised noise models, we checked the chromatic index $\alpha$ values used for the time-correlated noise components fixed at $0$, $2$ and $4$ for RN, DM and SV respectively. To do so, we used custom noise models for each pulsar and set $\alpha$ for each noise component as a hyperparameter to estimate its posterior distribution, as performed in recent PTA analyses \citep[e.g. ][]{gsr+21,cbp+22}. If the pulsar has two time-correlated components, we perform two analyses to evaluate each $\alpha$ independently while fixing the index of the other component to its relevant value ($0$ if RN and $2$ if DM). We use a uniform prior probability as $\mathcal{U} (-5,10)$. The medians and $95\%$ credible intervals of the posterior distributions for RN and DM chromatic indices (resp. $\alpha^{\mathrm{RN}}$ and $\alpha^{\mathrm{DM}}$) are shown in the first two columns for each dataset in Table (\ref{eptatable3}). In the following, we summarise our findings. 
\begin{itemize}
    \item For RN, we note that the chromatic indices of PSRs J1713+0747 and J1909$-$3744 are consistent with the expected value of $0$ with DR2new, while it is not the case with DR2full. We observe slightly negative values for PSRs J0030+0451, J0900$-$3144 and J1012+5307 with both datasets.
    \item For the DM chromatic indices, $18$ pulsars are consistent with the expected value for DM variations with both the DR2full+ and the DR2new. However, the high values for PSRs J1911+1347 and J2124$-$3358 with the DR2new warrant further investigation. 
    \item The only pulsar with a preferred SV model is PSR J1600$-$3053 with DR2new. The posterior distribution of the SV chromatic index for this pulsar is shown in the right panel of Figure \ref{J1600_timeserie}, with a median and $95\%$ credible intervals at $4.31^{+1.88}_{-1.27}$, nicely consistent with the expected value from scattering variations (cf. Section \ref{noisebudget}).
    \item For PSR J1022+1001, it is interesting that the trend in RN and DM indices swaps between the two datasets (that is, RN is shallower in DR2full+ and steeper in DR2new, while the opposite is the case for DM). To account for this, we stick to the standard noise model for this pulsar, as RN might prioritise the low frequencies, while DM also samples the high frequencies. The estimated indices for the DR2new show that the high-frequency signals are achromatic for this pulsar and the low-frequency components are consistent with the DM variations. While the former could correspond to profile instabilities \citep{ppb2020,lkl2015}, the latter is likely due to the presence of a variable solar wind \citep{tsb2021}.
\end{itemize}

For the second validity test, we evaluate a Bayes factor after including an achromatic noise component over the favoured models and therefore check if there is evidence for any remaining red noise. We use $30$ frequency components for this additional red noise in order to prioritise the low frequencies. As shown in column 5 of Table \ref{eptatable3}, we only find weak evidence (maximum of $2.3$ for PSR J1640+2224) for the inclusion of this component. These results reconfirm the robustness of the customised noise models that do not include an achromatic component.

\begin{table*}
\centering
\caption{Validation and performance statistics for the customised noise models with the ``DR2full+'' and ``DR2new''. For each dataset, the two first columns display the medians and $95\%$ confidence intervals of the chromatic index posterior distributions corresponding, respectively, to the RN and DM components. The values displayed in bold emphasise the numbers that are inconsistent with the expected values ($0$ for RN and $2$ for DM). The third columns show the Bayes factor for the customised model against the standard noise model (RN and DM with resp. $30$ and $100$ frequency bins). The last columns display the Bayes factor evaluations for an additional red noise component with $30$ frequency bins with the customised noise models.}
\label{eptatable3}
\begin{tabular*}{\textwidth}[c]{@{\extracolsep{\fill}}l|cccc|cccc}

\toprule
Pulsar &
\multicolumn{4}{c}{DR2full+} &
\multicolumn{4}{c}{DR2new}  \\
\cmidrule(lr){2-5} \cmidrule(lr){6-9}
& 
\multicolumn{1}{c}{$\alpha^{\mathrm{RN}}$} & 
\multicolumn{1}{c}{$\alpha^{\mathrm{DM}}$} &
\multicolumn{1}{c}{$\mathcal{B}^{\mathrm{cus}}_{\mathrm{std}}$} &
\multicolumn{1}{c|}{$\mathcal{B}^{\mathrm{cus+RN30}}_{\mathrm{cus}}$} &
\multicolumn{1}{c}{$\alpha^{\mathrm{RN}}$} & 
\multicolumn{1}{c}{$\alpha^{\mathrm{DM}}$} &
\multicolumn{1}{c}{$\mathcal{B}^{\mathrm{cus}}_{\mathrm{std}}$} &
\multicolumn{1}{c}{$\mathcal{B}^{\mathrm{cus+RN30}}_{\mathrm{cus}}$}\\
\midrule

J0030+0451 & $-1.23^{+1.63}_{-1.08}$ & X & $1.8$ & $1.0$ & $-4.03^{+5.54}_{-0.91}$ & X & $1.6$ & $1.1$ \\[3pt]
J0613$-$0200 & $0.85^{+0.44}_{-0.87}$ & $\mathbf{2.93^{+0.38}_{-0.43}}$ & $79.9$ & $1.0$ & X & $4.31^{+5.11}_{-4.39}$ & $3.8$ & $1.7$ \\[3pt]
J0751+1807 & X & $1.59^{+0.91}_{-0.68}$ & $2.8$ & $1.9$ & X & $1.34^{+1.26}_{-1.09}$ & $1.0$ & $1.1$ \\[3pt]
J0900$-$3144 & $-0.85^{+1.54}_{-1.55}$ & $1.35^{+7.19}_{-0.64}$ & $\geq 10^3$ & $1.2$ & $-0.88^{+1.84}_{-1.77}$ & $4.77^{+4.98}_{-3.88}$ & $287.8$ & $1.5$ \\[3pt]
J1012+5307 & $\mathbf{-0.65^{+0.46}_{-0.41}}$ & $2.04^{+0.66}_{-0.46}$ & $\geq 10^3$ & $2.0$ & $\mathbf{-0.56^{+0.45}_{-0.43}}$ & $2.10^{+5.16}_{-1.04}$ & $\geq 10^3$ & $1.8$\\[3pt]
J1022+1001 & $1.02^{+2.28}_{-3.13}$ & $\mathbf{1.54^{+0.41}_{-0.37}}$ & X & $1.0$ & $3.04^{+6.34}_{-7.12}$ & $\mathbf{0.28^{+1.48}_{-1.07}}$ & X & $1.6$\\[3pt]
J1024$-$0719 & X & $1.42^{+2.18}_{-1.18}$ & $1.2$ & $1.2$ & X & $1.92^{+2.44}_{-1.53}$ & $3.0$ & $1.6$ \\[3pt]
J1455$-$3330 & $0.58^{+5.10}_{-3.60}$ & X & $1.2$ & $1.0$ & $1.39^{+6.18}_{-4.71}$ & X & $1.0$ & $1.2$\\[3pt]
J1600$-$3053 & $1.95^{+0.27}_{-3.21}$ & $2.15^{+0.41}_{-0.30}$ & $\geq 10^3$ & $1.2$ & X & $2.16^{+0.31}_{-0.47}$ & $\geq 10^3$ & $1.2$\\[3pt]
J1640+2224 & X & $2.50^{+5.74}_{-2.34}$ & $2.6$ & $1.4$ & X & $2.36^{+5.33}_{-2.20}$ & $3.4$ & $2.3$\\[3pt]
J1713+0747 & $\mathbf{-0.85^{+0.63}_{-0.78}}$ & $2.02^{+0.20}_{-0.21}$ & $\geq 10^3$& $1.2$ & $0.25^{+0.74}_{-1.96}$ & $1.76^{+0.47}_{-0.69}$ & $\geq 10^3$ & $1.1$ \\[3pt]
J1730$-$2304 & X & $2.85^{+2.32}_{-2.04}$ & $6.6$ & $1.7$ & X & $3.11^{+3.15}_{-2.42}$ & $3.4$ & $1.8$\\[3pt]
J1738+0333 & $0.73^{+2.55}_{-2.29}$ & X & $1.8$ & $1.0$ & X & $1.53^{+2.65}_{-1.42}$ & $1.3$ & $1.0$\\[3pt]
J1744$-$1134 & $\mathbf{1.01^{+0.70}_{-0.71}}$ & $1.85^{+1.68}_{-1.28}$ & $\geq 10^3$ & $0.8$ & X & $1.85^{+1.68}_{-1.28}$ & $2.4$ & $2.0$\\[3pt]
J1751$-$2857 & X & $4.60^{+3.87}_{-3.58}$ & $1.9$ & $1.3$ & X & $5.79^{+3.90}_{-4.66}$ & $3.0$ & $1.6$ \\[3pt]
J1801$-$1417 & X & $3.64^{+2.03}_{-1.73}$ & $27.1$ & $1.8$ & X & $3.77^{+2.06}_{-1.74}$ & $8.1$ & $1.6$\\[3pt]
J1804$-$2717 & X & $2.66^{+4.62}_{-2.36}$ & $6.3$ & $1.7$ & X & $2.99^{+5.87}_{-2.70}$ & $6.2$ & $1.9$\\[3pt]
J1843$-$1113 & X & $1.66^{+0.97}_{-0.64}$ & $3.3$ & $1.9$ & X & $1.89^{+0.36}_{-0.37}$ & $2.1$ & $2.0$\\[3pt]
J1857+0943 & X & $2.19^{+0.44}_{-0.57}$ & $8.1$ & $2.3$ & X & $2.80^{+1.99}_{-1.52}$ & $2.1$ & $2.1$\\[3pt]
J1909$-$3744 & $0.60^{+0.35}_{-0.87}$ & $\mathbf{2.31^{+0.14}_{-0.14}}$ & $\geq 10^3$ & $1.0$ & $0.35^{+0.44}_{-0.76}$ & $\mathbf{3.10^{+0.88}_{-0.74}}$ & $3.0$ & $1.1$\\[3pt]
J1910+1256 & X & $3.75^{+3.31}_{-3.07}$ & $2.2$ & $1.8$ & X & $3.35^{+2.78}_{-2.57}$ & $1.7$ & $1.7$\\[3pt]
J1911+1347 & X & $3.50^{+1.89}_{-1.57}$ & $2.7$ & $2.0$ & X & $\mathbf{7.60^{+2.28}_{-4.21}}$ & $1.8$ & $1.8$\\[3pt]
J1918$-$0642 & X & $2.57^{+1.69}_{-1.74}$ & $2.1$ & $2.0$ & X & $2.70^{+1.87}_{-1.75}$ & $1.9$ & $2.0$\\[3pt]
J2124$-$3358 & X & $1.60^{+0.84}_{-1.15}$ & $2.7$ & $1.6$ & X & $\mathbf{8.12^{+1.80}_{-4.48}}$ & $4.9$ & $2.0$\\[3pt]
J2322+2057 & X & X & $4.1$ & $2.0$ & X & X & $3.7$ & $1.9$ \\[3pt]
\bottomrule
\end{tabular*}%
\end{table*}%

\subsection{Consistency among different softwares}
\label{ent_vs_tempones}

In Table \ref{SPNA_SPNTA_Tension}, we present the tension in $\sigma$ (see Section \ref{InPTA} for a description of the tension metric) for red and DM noise models estimated using two Bayesian timing packages, \texttt{enterprise} and \texttt{temponest} for the 25 EPTA MSPs. We find that the noise models for all the pulsars are very consistent.

\begin{table}
    \caption{Tension metrics for the red and DM noise models estimated using \texttt{enterprise} and \texttt{temponest}.}
    \centering
    \begin{tabular}{l|l|l|l}
    \hline
        \textbf{Pulsar} & \textbf{Model} & \textbf{RN Tension} & \textbf{DM Tension} \\ \hline
        J0030+0451 & RN & 0.02 & X \\ 
        J0613$-$0200 & DM+RN & 0.40 & 0.07 \\ 
        J0751+1807 & DM & X & 0.07 \\ 
        J0900$-$3144 & DM+RN & 0.04 & 0.02 \\ 
        J1012+5307 & DM+RN & 0.06 & 0.66 \\ 
        J1022+1001 & DM+RN & 0.01 & 0.23 \\ 
        J1024$-$0719 & DM & X & 0.16 \\ 
        J1455$-$3330 & RN & 0.02 & X \\ 
        J1600$-$3053 & DM & X & 0.03 \\ 
        J1640+2224 & DM & X & 0.16\\ 
        J1713+0747 & DM+RN & 0.01 & 0.11 \\ 
        J1730$-$2304 & DM & X & 0.001 \\ 
        J1738+0333 & RN & 0.001 & X \\ 
        J1744$-$1134 & DM+RN & 0.06 & 0.08 \\ 
        J1751$-$2857 & DM & X & 0.01 \\ 
        J1801$-$1417 & DM & X & 0.01 \\ 
        J1804$-$2717  & DM & X & 0.01 \\ 
        J1857+0943 & DM & X & 0.21 \\ 
        J1909$-$3744 & DM+RN & 0.03 & 0.04 \\ 
        J1910+1256 & DM & X & 0.001 \\ 
        J1911+1347 & DM & X & 0.001 \\ 
        J1918$-$0642 & DM & X & 0.02 \\ 
        J2124$-$3358 & DM & X & 0.01 \\ 
        J2322+2057 & X & X & X \\ \hline
    \end{tabular}
    \label{SPNA_SPNTA_Tension}
\end{table}

\subsection{Marginalisation of the Timing Model}
\label{marginalisation}
Preliminary analysis of DR2full using \texttt{temponest} showed that the posteriors of the noise model parameters varied significantly depending on whether the timing model was marginalised. This was particularly surprising given that \citet{lhf+14} demonstrated that neither the analytic marginalisation nor linearisation made a significant difference to any parameters. Further investigation revealed that changes in the implementation of \texttt{temponest} in 2017 \footnote{commit hash: \texttt{e745752}} introduced a critical flaw when not marginalising over the timing model whilst marginalising over the arbitrary jumps between instruments; the typical method of operation when not marginalising over the timing model.
This error disabled fitting for many jumps, leading to incorrect noise model inferences. Once resolved, we no longer find any discrepancy between noise model parameters when marginalising over the timing model, and find that the linearised timing model is sufficient for further analysis once any highly non-linear binary parameters have been solved.

\subsection{Simulations}
We have implemented a toolkit to generate repeatable simulations of EPTA datasets to validate noise models.
Simulations are generated using the \texttt{toasim} framework that is distributed with \texttt{tempo2}.
In brief, the method is to generate a set of `idealised' TOAs that produce zero residual with respect to a seed pulsar parameter file and then add perturbations for each of the simulated noise processes.
For an idealised TOA $t_i$, the final simulated TOA is given by
\begin{equation}
    t_{\mathrm{final},i} = t_i + y_\mathrm{red}(t_i)+ y_\mathrm{dm}(t_i)+ y_\mathrm{scat}(t_i)+ y_\mathrm{white}(t_i),
\end{equation}
where each of the perturbations $y$ is discussed in the following sections.
In principle, this can only be solved iteratively, but in practise we can neglect second-order corrections as long as the perturbations are small.
Therefore, we typically subtract a quadratic from each $y(t)$, which significantly reduces the general amplitude of the perturbations, especially for steep red noise processes.
Removing a quadratic in this way does not affect the noise models, but it does mean that the mean pulsar spin frequency and frequency derivative are unchanged from realisation to realisation.
This is hence equivalent to fitting for \texttt{F0} and \texttt{F1} on each dataset, and small variations in these parameters are largely uninteresting, and in many senses arbitrary for a simulation, this does not affect the overall usefulness of the simulations.

Significant improvements have been implemented to the \texttt{toasim} code as part of this work, which were required to ensure that the definitions of the model parameters are consistent between this work and the simulation code.
The simulation code has been adapted to be consistent with the \texttt{tempo2} and \texttt{temponest} definitions of the model parameters, although these are generally trivially related to other definitions.

\subsubsection{Achromatic red noise}
The \texttt{toasim} framework provides a method for simulating power-law red-noise processes by means of the inverse Fourier transform.
This first computes the noise process over a grid of 4096 equally spaced values over $T_\mathrm{span}$, extended by a factor of $n=100$ to avoid periodic boundary effects of the Fourier transform.
The spectrum is computed for $N=1024n$ frequencies with Hermitian symmetry, and Fourier transformed to give a time series:
\begin{equation}
    r_k = \sum\limits_{j=-N/2}^{N/2} R_j (a_j + ib_j)\exp{(2\pi \sqrt{-1}jk/N)},
\end{equation}
with $a_j$ and $b_j$ random variables drawn from $\mathcal{N}(0,1)$.
Note that the requirement for Hermitian symmetry means that only $512$ independent random values are needed for each of $a_j$ and $b_j$. This is largely analogous to Equation \ref{noise_equation} for the noise modelling except that the Fourier transform includes both positive and negative frequencies and therefore the amplitudes must be scaled down by a factor of two, relative to the noise models in Equation \ref{noise_PSD}.
Therefore, the mean amplitude at each frequency $R_j$ is given by
\begin{equation} 
\label{simulation_red_amp}
    R_j =\frac{1}{2}\sqrt{\frac{A_\mathrm{red}^2}{12\pi^2}\frac{\mathrm{s_\mathrm{yr}}^3}{nT_\mathrm{span}}\left(\frac{f_j}{f_\mathrm{yr}}\right)^{-\gamma_\mathrm{red}}},
\end{equation}
where $f_\mathrm{yr} = 1 \mathrm{yr}^{-1}$, and $\mathrm{s_\mathrm{yr}} = 31557600\,\mathrm{s\,yr}^{-1}$ converts years to seconds to give a perturbation in seconds for $A_\mathrm{red}$ in $\mathrm{yr}^{3/2}$ and $T_\mathrm{span}$ in s.
The factor of $1/2$ corrects from the one-sided PSD used by the noise models to the two-sided PSD needed for the Fourier transform.
The final perturbation for the achromatic red noise for TOA $t_i$ is computed by linear interpolation of $r_k$,
\begin{equation}
\label{simulation_red_interpolate}
    y_\mathrm{red}(t_i) = r_k + (t_i-t_k)\frac{r_{k+1}-r_k}{t_{k+1}-t_k},
\end{equation}
for $t_k < t_i < t_{k+1}$.
This is implemented by the \texttt{addRedNoise} \texttt{tempo2} plugin.

\subsubsection{DM Variations}
Perturbations due to DM variations are similarly computed by the inverse Fourier transform using largely the same method as for the achromatic noise, except that we model a DM time series before converting to a TOA perturbation as a last step.
The mean amplitude at each frequency, $D_{\rm j}$ is given by
\begin{equation} 
\label{simulation_dm_amp}
    D_j =\frac{1}{2}\sqrt{A_\mathrm{DM}^2\frac{\mathrm{s_\mathrm{yr}}^3}{nT_\mathrm{span}}\left(\frac{f_j}{f_\mathrm{yr}}\right)^{-\gamma_\mathrm{DM}}},
\end{equation}
with $T_\mathrm{span}$ in seconds, $A_\mathrm{DM}$ in \texttt{temponest} units of $\mathrm{cm}^{-3}\mathrm{pc}\,\mathrm{yr}^{3/2}\mathrm{s}^{-1}$.

The final perturbation for a given TOA is then calculated from $d(t)$, a function that linearly interpolates $d_{\rm k}$ in a similar way to Equation \ref{simulation_red_interpolate}, and
\begin{equation}
    y_\mathrm{DM}(t_i) = \frac{d(t_i)}{\kappa_\mathrm{DM} \nu_k^2},
\end{equation}
where $\nu_k$ is the observing frequency of the TOA and $\kappa_\mathrm{DM} = 2.41\times10^{-4} \mathrm{cm^{-3}pc\,MHz^2s^{-1}}$ is the DM constant.
This is implemented by the \texttt{addDmVar} plugin in \texttt{tempo2}.

\subsubsection{Scattering Variations}
Scattering variations are implemented very similarly to the DM variations.
We create a scattering time series $s_{\rm k}$ using the Fourier transform of $n$ Hermitian complex values given by $S_j(a_j+ib_j)$, where
\begin{equation} 
\label{simulation_scat_amp}
    S_j =\frac{1}{2}\sqrt{\frac{A_\mathrm{scat}^2}{12\pi^2}\frac{\mathrm{s_\mathrm{yr}}^3}{nT_\mathrm{span}}\left(\frac{f_j}{f_\mathrm{yr}}\right)^{-\gamma_\mathrm{scat}}}.
\end{equation}
Then the perturbation associated with the scattering is computed from a linear interpolation function $s(t)$ to give
\begin{equation}
    y_\mathrm{scat}(t_i) = s(t_i) \left(\frac{\nu}{\nu_\mathrm{ref}}\right)^{-\alpha_\mathrm{scat}},
\end{equation}
where we use $\alpha_\mathrm{scat} =4 $ for scattering variations, and $\nu_\mathrm{ref}=1400\,\mathrm{MHz}$, chosen to be consistent with the implementation in \texttt{temponest} and \texttt{enterprise}.
This is implemented by the \texttt{addChromVar} \texttt{tempo2} plugin.

\subsubsection{White Noise}
White noise is added to the TOA with uncertainty $\sigma_i$ after applying the EFAC and EQUAD parameters (see Equation \ref{efacequad}) by drawing $y_\mathrm{white}(t_i)$ from $\mathcal{N}(0,\sigma_i^2)$.

\subsubsection{Creating the realisations}
The simulation parameters can be given as single values or a uniform prior range from which to draw, so that a different set of parameters is used for each realisation of the simulation.
When the parameters are drawn randomly from the prior, we record the sets of parameters used for later comparison with the results.

\subsection{P-P plots}
One tool to validate Bayesian analysis tools is by studying posterior quantiles from simulated datasets \citep{cgr12}.
Our method largely follows that implemented in \texttt{bilby} \citep{ahl+19}, where we simulate a large number of datasets, $j$, with `true' model parameters $\boldsymbol{\Theta}_j$ drawn from a prior distribution $p({\boldsymbol \Theta})$.
We then use our Bayesian software, in this case \texttt{temponest}, to generate samples, $\theta_{i,j}$ of the posterior distribution of the parameters given the prior distribution (see Table \ref{pp_priors}), for each of the simulated datasets, following the same approach as for the real data.
We can then, for each parameter $i$, and each simulated dataset $j$, compute the quantile $q_{{i,j}}$, within the 1-d posterior that the `true' value of the parameter lies, i.e. the weighted fraction of samples for which $\theta_{i,j} < \Theta_{i,j}$. It can be shown that the $q_{i,j}$ should be uniformly distributed between 0 and 1 \citep{cgr12}.
We can visualise this by plotting the cumulative distribution of $q_{i,j}$ over all $j$, yielding a so-called `P-P plot'.
In principle, this should follow a straight line, but even with perfect algorithms, there will be deviations caused by the finite number of simulations, which we can estimate with confidence contours taken from the binomial distribution.

\begin{table}
    \caption{Uniform prior bounds for the P-P simulations for J1600$-$3053. A separate EFAC and EQUAD is drawn for each \texttt{-group} flag.}
    \label{pp_priors}
    \centering
    \begin{tabular}{r|l|l}
    \toprule
    Parameter & Low & High \\
    \midrule
        $\log{(A_\mathrm{red})}$ & $-15.0$&$-13.4$ \\
        $\gamma_\mathrm{red}$ & $2.0$&$6.0$ \\
        $\log{(A_\mathrm{dm})}$  & $-13.0$&$-11.0$ \\
        $\gamma_\mathrm{dm}$ & $1.6$&$3.2$ \\ $\log{(A_\mathrm{scat})}$ & $-15.0$&$-13.2$ \\
        $\gamma_\mathrm{scat}$ & $1.1$&$2.2$ \\
        EFAC & $0.5$&$2.0$ \\
        $\log_{10}$ EQUAD & $-8$&$-6.5$\\
        \bottomrule

    \end{tabular}
    
\end{table}

It is important that the prior distribution matches between the simulated data and the Bayesian search, particularly for any parameters that may not be constrained on one or both sides and hence be bounded by the prior.
This is quite common in the pulsar datasets, as power-law red-noise processes below the detection threshold will have log amplitude unbounded on the low side and the spectral slope may simply return the prior.
If the prior does not match the simulation and posterior generation, this will inevitably lead to a perceived error in the posterior computation for such cases.
We choose a prior range such that most simulations have noise processes that are largely representative of what we estimated in the real data, and wide enough that the posterior of the real data would not be heavily constrained by the choice of prior.

\begin{figure*}
    \centering
    \begin{tabular}{cc}
    \includegraphics[width=0.5\textwidth]{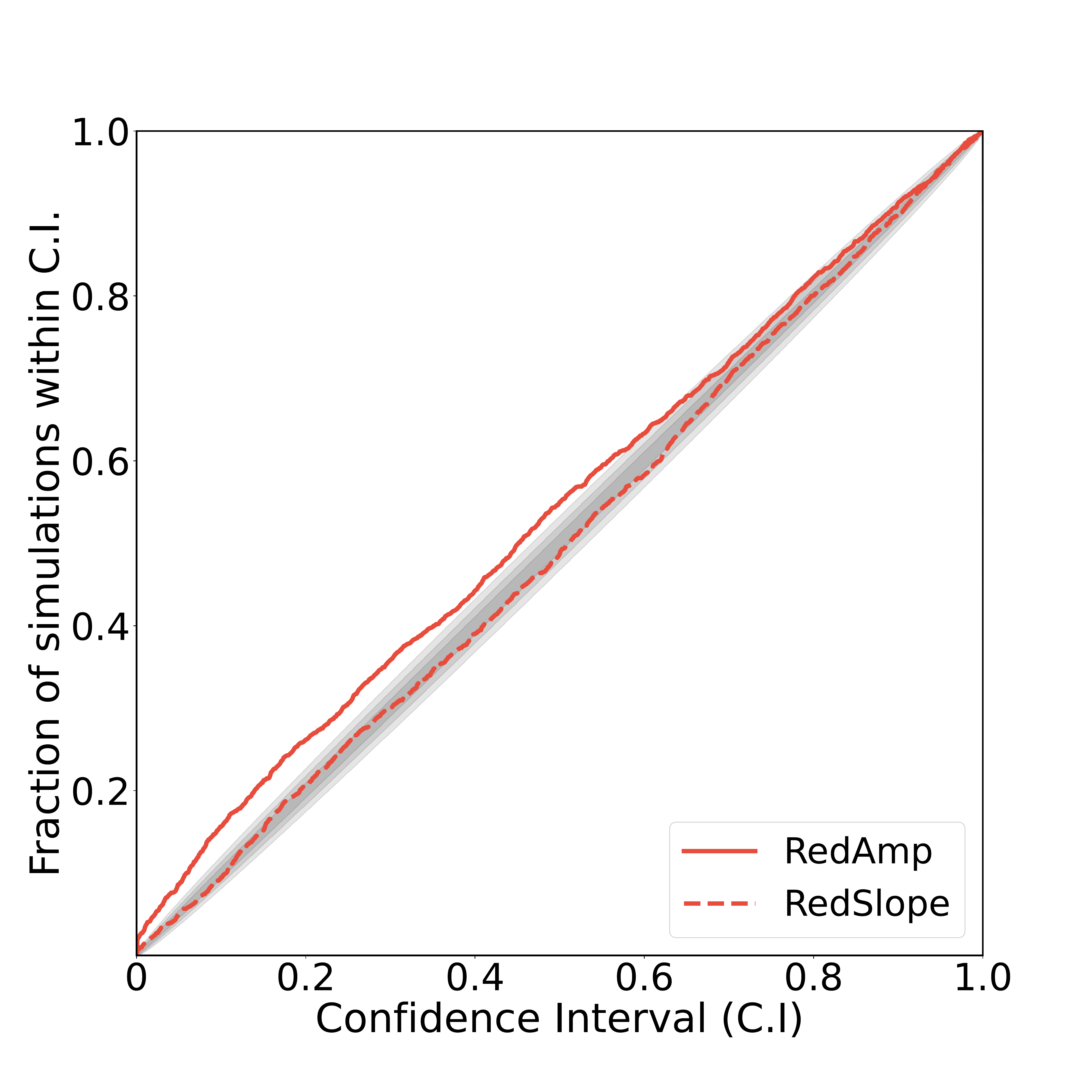}
    \includegraphics[width=0.5\textwidth]{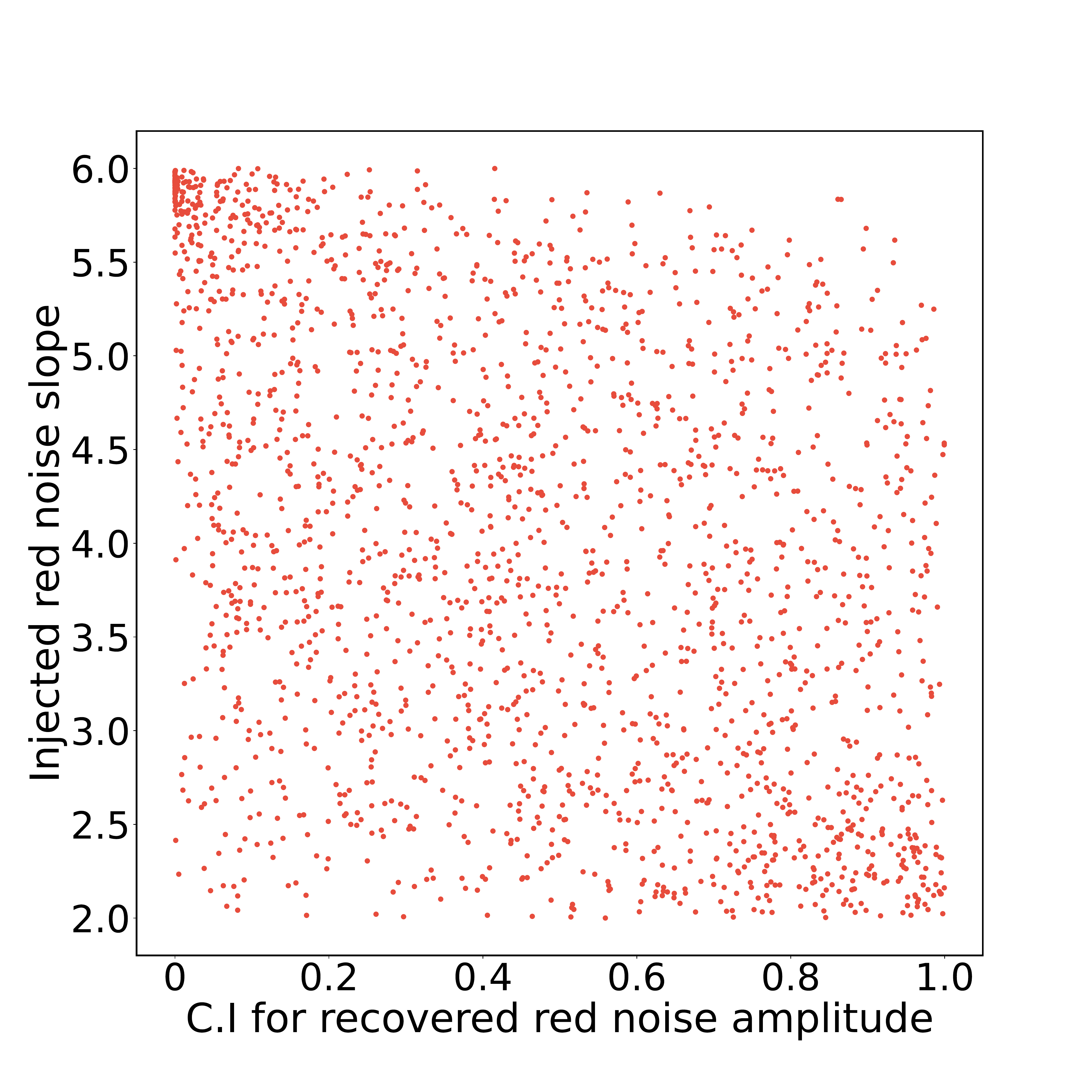}\\
    \includegraphics[width=0.5\textwidth]{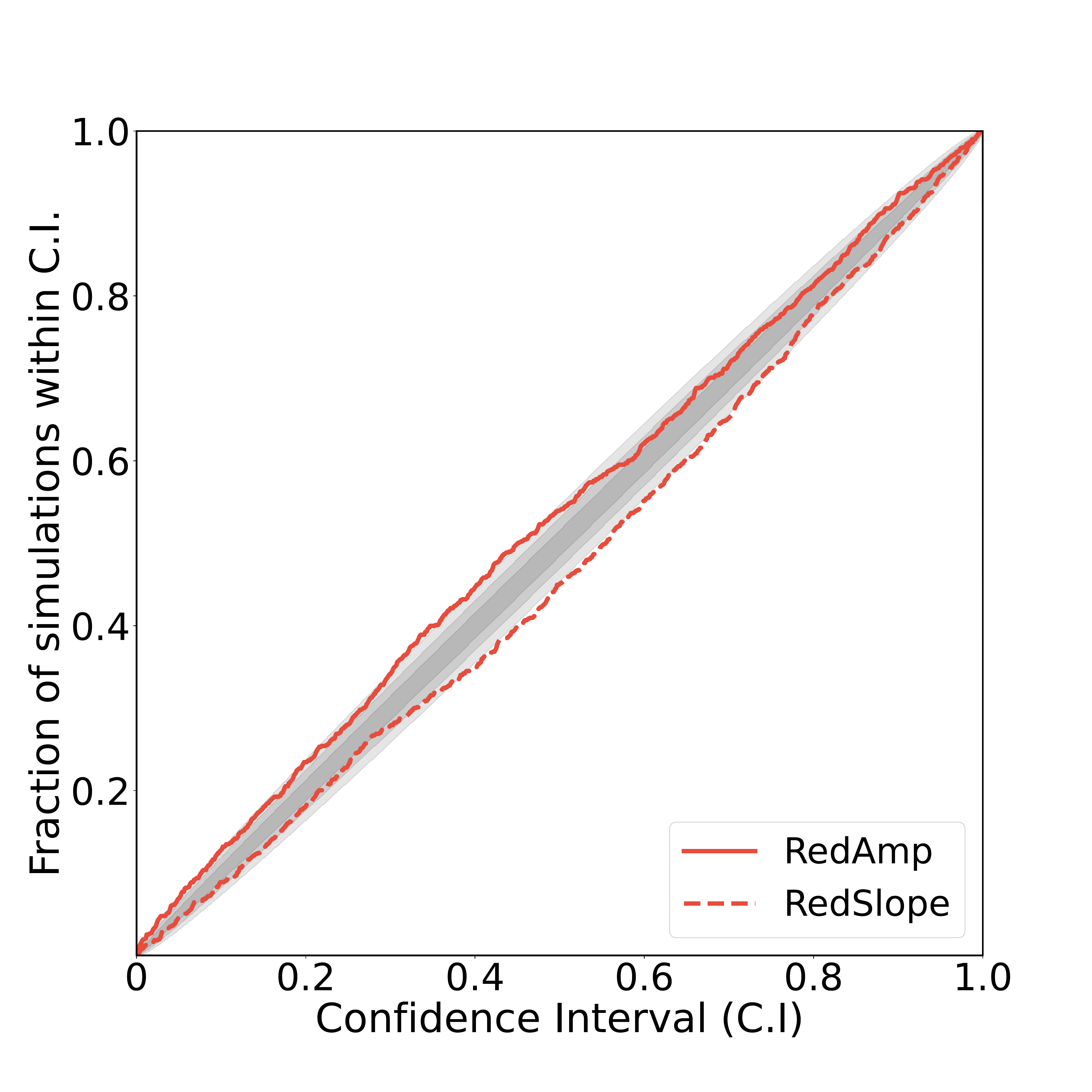}
    \includegraphics[width=0.5\textwidth]{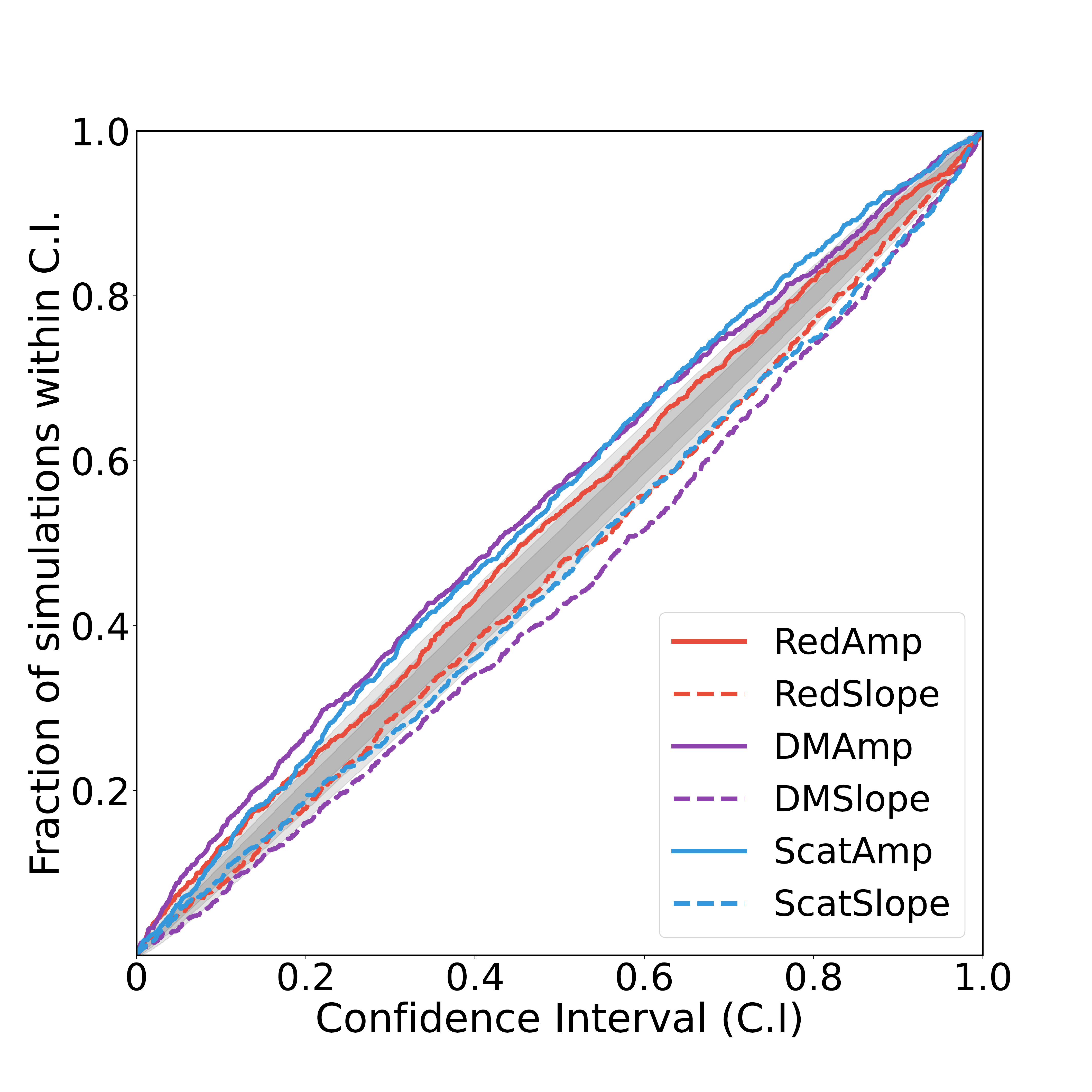}
    \end{tabular}
    \caption{(From top left to bottom right) (a) P-P plot for $\sim$ 2000 simulations with only chromatic noise, injected at much lower than $1/T_\mathrm{span}$. (b) The confidence interval of the recovered achromatic amplitude for various levels of injected red noise slope for the simulations is shown in (a). (c) P-P plot for $\sim$ 1000 simulations with only chromatic noise periodic on $1/T_\mathrm{span}$. (d) P-P plot for $\sim$ 1000 simulations with chromatic, achromatic and scattering variations, periodic on $1/T_\mathrm{span}$.}
    \label{pplot}
\end{figure*}

\textbf{Results:} For simplicity, we first consider the simulation of only achromatic red noise. The P-P plot, shown in Figure \ref{pplot} (a), indicates generally good consistency in the posterior of $\gamma_\mathrm{red}$, but there is an excess at low confidence intervals for $\log{(A_\mathrm{red})}$.
The investigation of individual results shows a correlation between the $\log{(A_\mathrm{red})}$ confidence interval and the injected $\gamma_\mathrm{red}$, visualised in Figure~\ref{pplot} (b).
We believe that the observed correlation suggests that the excess results are driven by cases where $\gamma_\mathrm{red}$ is large, and hence we suspect that they are cases where the
red noise has such a steep spectrum that it is no longer able to be expressed in a Fourier basis with lowest frequency at $1/T_\mathrm{span}$.

To understand this, we recall that the Fourier-basis Gaussian process models used in this work, and widely used by other similar projects are naturally periodic on $T_\mathrm{span}$ due to the nature of the Fourier basis.
This use of a Fourier basis vastly speeds up the computation compared to directly evaluating the covariance matrix for these large datasets, but it requires that the noise process can be well approximated by a periodic function.
It is generally assumed that fitting for the spin frequency (\texttt{F0}) and spin frequency derivatives (\texttt{F1}) parameters absorbs the low frequencies sufficiently that the periodic model is acceptable, and previous analyses have suggested that the posteriors remain largely unchanged when trying to absorb noise below $1/T_\mathrm{span}$ even for pulsars with $\gamma_\mathrm{red}\sim 6$ \citep{lsc+16}.
However, the P-P plot is very sensitive to even small biases, and hence quite likely that the difference may not be noticeable (or have strong Bayesian evidence) for a single analysis.
Recently, it has also been seen that Fourier-basis Gaussian processes models periodic on $T_\mathrm{span}$ struggle to recover the spin frequency second derivative (\texttt{F2}) when $\gamma_\mathrm{red}$ is greater than $\sim 4$ \citep{kn23}, and this implies that fitting for \texttt{F0} and \texttt{F1} may not be sufficient to allow safe use of the $1/T_\mathrm{span}$ Fourier basis model when the noise process has a very steep spectral exponent.

In order to test our hypothesis, we repeat the analsis whilst forcing the injected red noise to be periodic on $T_\mathrm{span}$, with results shown in Figure \ref{pplot} (c).
The curve now goes smoothly to zero at 0, and there is no longer any correlation between the $\log{(A_\mathrm{red})}$ confidence interval and the injected $\gamma_\mathrm{red}$.
However, the results are still not consistent with a one-to-one relationship (with probability $< 0.5\%$), in particular, we find that $\log{(A_\mathrm{red})}$ is slightly biased high and $\gamma_\mathrm{red}$ biased slightly low.
We find that a correction of $\gamma_\mathrm{red}$ by only  $\sim2\%$ is needed to fully correct the bias, a correction much smaller than the typical uncertainty of the measurement.
The slope and amplitude are highly correlated, and indeed scaling $\log{(A_\mathrm{red})}$ to a longer time span closer to where the signal can be measured is sufficient to remove the apparent bias here.
Our hypothesis is that there is an intrinsic bias for slightly flatter spectral slopes, which then manifests itself as a corresponding bias in the amplitude at 1yr$^{-1}$.

We repeat this with simulations that include achromatic noise red noise, excess white noise, DM variations, and scattering variations.
The resulting P-P plots are very similar to the simulation with achromatic red noise only.
Figure~\ref{pplot} (d) shows the P-P plot when the noise is periodic on $1/T_\mathrm{span}$, showing a similarly small but significant bias in the posterior parameters for all three parameters. The similarity between all three components of the noise model to that obtained when only injecting achromatic noise suggests that the bias is due to something fundamental in the Fourier domain Gaussian process modelling of the noise, rather than being driven by the choice of the dataset.

Overall, the P-P plots are encouraging and show that the recovered posteriors are largely correct, but in almost all cases, we find subtle deviations from the correct posterior distribution.
In agreement with section \ref{ent_vs_tempones}, we find that the analysis does not change significantly if using \texttt{enterprise} to compute the noise models rather than \texttt{temponest}.
The P-P plot methodology is sensitive to very subtle errors in the posterior, including biases much smaller than the uncertainty on the results, so although the results are not perfect, we do not feel that it invalidates the results we present.
Nevertheless, we feel that it is important that these tests are repeated on a wider scale across the IPTA, to fully understand the characteristics of the posteriors produced by our noise modelling code and increase confidence in the interpretation of the posterior distributions for the pulsar noise models and the GWB detection parameters, particularly for any pulsars with steep spectrum red noise.

\section{Conclusions}
\label{conclusions}
The noise models presented here are our current best estimates for stochastic noise in the 25-pulsar EPTA DR2 dataset.
With the inclusion of wide-band receivers and additional low-frequency data from the InPTA collaboration, all but four pulsars show evidence for chromatic noise, specifically in the form of DM variations. However, several of these pulsars seem to have power-law noise with a much flatter spectrum than expected from DM variations in the turbulent IISM. This may be attributed to additional high-frequency terms, but it may also suggest other processes leaking into the model for DM variations. Compared to the EPTA DR1 and InPTA datasets, we find that the choice of preferred noise model can change, including a couple of cases where chromatic noise that was ruled out in EPTA DR1 becomes detected in EPTA DR2. 

We feel that this is evidence that the assumptions typically made in noise modelling are not strictly true for all pulsars. Either there are additional processes that need to be considered, such as noise that is not purely modelled by a power law, or the spectral properties of the noise vary over time, or there are systematic instrumental effects that affect particular parts of the dataset. In reality, none of these assumptions are likely completely true, and the improvements in data quality and time span of the EPTA DR2 data particularly highlight these subtle effects.

The fact that the best noise models can depend on the frequency and time coverage of the dataset implies that great care is needed when comparing the noise models across different PTAs, and hence we suggest that a thorough investigation into the time and frequency stationary of the pulsar noise models, as well as exploration of additional noise terms, is best undertaken within the IPTA framework, which necessarily has better time and frequency coverage than any individual PTA.

We have also tested the estimation of hyperparameters from the noise model in \texttt{temponest} and \texttt{enterprise} and find that, as might be expected, they are highly consistent in results in real and simulated data. We also demonstrate that previously observed differences in model parameters when marginalising over the timing model parameters in \texttt{temponest} were due to a software bug and that marginalising over the timing model is entirely sufficient when trying to estimate the noise model hyperparameters.

Furthermore, we performed a `P-P plot' test on our noise model hyperparameter estimation using simulated data and found that there may be two subtle biases in the results. For `realistic' simulations where there is power at timescales longer than the observing span, a bias in the estimation of the red noise amplitude is observed when the spectral slope of the noise is greater than $\sim 4$. We further saw that even if we simulate noise that has the same periodic nature as the noise model, there is a very small bias ($\sim 2\%)$ in the recovered spectral slope. Although these tests require large numbers of trials and, hence, large computing costs, we encourage further testing of this nature within the IPTA framework to attempt to better understand the origin and any possible effect of these biases.

It is important to note that although we find several areas of investigation for a better understanding of the pulsar noise processes, the custom noise models presented here model the noise in the EPTA dataset extremely well, and the majority of the pulsar noise models are consistent between all datasets that we have compared. We also demonstrate that our customised noise models improve our overall sensitivity to GWB signals over the `standard' models for both the complete DR2full+ dataset as well as when focussing only on the new purpose-built EPTA instrumentation in DR2new.

\begin{acknowledgements}
The European Pulsar Timing Array (EPTA) is a collaboration between
European and partner institutes, namely ASTRON (NL), INAF/Osservatorio
di Cagliari (IT), Max-Planck-Institut f\"{u}r Radioastronomie (GER),
Nan\c{c}ay/Paris Observatory (FRA), the University of Manchester (UK),
the University of Birmingham (UK), the University of East Anglia (UK),
the University of Bielefeld (GER), the University of Paris (FRA), the
University of Milan-Bicocca (IT), the Foundation for Research and 
Technology, Hellas (GR), and Peking University (CHN), with the
aim to provide high-precision pulsar timing to work towards the direct
detection of low-frequency gravitational waves. An Advanced Grant of
the European Research Council allowed to implement the Large European Array
for Pulsars (LEAP) under Grant Agreement Number 227947 (PI M. Kramer). 
The EPTA is part of the
International Pulsar Timing Array (IPTA); we thank our
IPTA colleagues for their support and help with this paper and the external Detection Committee members for their work on the Detection Checklist.

Part of this work is based on observations with the 100-m telescope of
the Max-Planck-Institut f\"{u}r Radioastronomie (MPIfR) at Effelsberg
in Germany. Pulsar research at the Jodrell Bank Centre for
Astrophysics and the observations using the Lovell Telescope are
supported by a Consolidated Grant (ST/T000414/1) from the UK's Science
and Technology Facilities Council (STFC). ICN is also supported by the
STFC doctoral training grant ST/T506291/1. The Nan{\c c}ay radio
Observatory is operated by the Paris Observatory, associated with the
French Centre National de la Recherche Scientifique (CNRS), and
partially supported by the Region Centre in France. We acknowledge
financial support from ``Programme National de Cosmologie and
Galaxies'' (PNCG), and ``Programme National Hautes Energies'' (PNHE)
funded by CNRS/INSU-IN2P3-INP, CEA and CNES, France. We acknowledge
financial support from Agence Nationale de la Recherche
(ANR-18-CE31-0015), France. The Westerbork Synthesis Radio Telescope
is operated by the Netherlands Institute for Radio Astronomy (ASTRON)
with support from the Netherlands Foundation for Scientific Research
(NWO). The Sardinia Radio Telescope (SRT) is funded by the Department
of University and Research (MIUR), the Italian Space Agency (ASI), and
the Autonomous Region of Sardinia (RAS) and is operated as a National
Facility by the National Institute for Astrophysics (INAF).

The work is supported by the National SKA programme of China
(2020SKA0120100), Max-Planck Partner Group, NSFC 11690024, CAS
Cultivation Project for FAST Scientific. This work is also supported
as part of the ``LEGACY'' MPG-CAS collaboration on low-frequency
gravitational wave astronomy. JA acknowledges support from the
European Commission (Grant Agreement number: 101094354). JA and SCha 
were partially supported by the Stavros
Niarchos Foundation (SNF) and the Hellenic Foundation for Research and
Innovation (H.F.R.I.) under the 2nd Call of the ``Science and Society --
Action Always strive for excellence -- Theodoros Papazoglou''
(Project Number: 01431). AC acknowledges support from the Paris
\^{I}le-de-France Region. AC, AF, ASe, ASa, EB, DI, GMS, MBo acknowledge
financial support provided under the European Union's H2020 ERC
Consolidator Grant ``Binary Massive Black Hole Astrophysics'' (B
Massive, Grant Agreement: 818691). GD, KLi, RK and MK acknowledge support
from European Research Council (ERC) Synergy Grant ``BlackHoleCam'', 
Grant Agreement Number 610058. This work is supported by the ERC 
Advanced Grant ``LEAP'', Grant Agreement Number 227947 (PI M. Kramer). 
AV and PRB are supported by the UK's Science
and Technology Facilities Council (STFC; grant ST/W000946/1). AV also acknowledges
the support of the Royal Society and Wolfson Foundation. JPWV acknowledges
support by the Deutsche Forschungsgemeinschaft (DFG) through thew
Heisenberg programme (Project No. 433075039) and by the NSF through
AccelNet award \#2114721. NKP is funded by the Deutsche
Forschungsgemeinschaft (DFG, German Research Foundation) --
Projektnummer PO 2758/1--1, through the Walter--Benjamin
programme. ASa thanks the Alexander von Humboldt foundation in
Germany for a Humboldt fellowship for postdoctoral researchers. APo, DP
and MBu acknowledge support from the research grant “iPeska”
(P.I. Andrea Possenti) funded under the INAF national call
Prin-SKA/CTA approved with the Presidential Decree 70/2016
(Italy). RNC acknowledges financial support from the Special Account
for Research Funds of the Hellenic Open University (ELKE-HOU) under
the research programme ``GRAVPUL'' (grant agreement 319/10-10-2022).
EvdW, CGB and GHJ acknowledge support from the Dutch National Science
Agenda, NWA Startimpuls – 400.17.608.
BG is supported by the Italian Ministry of Education, University and 
Research within the PRIN 2017 Research Program Framework, n. 2017SYRTCN. LS acknowledges the use of the HPC system Cobra at the Max Planck Computing and Data Facility.

\ifnum\wm>1 The Indian Pulsar Timing Array (InPTA) is an Indo-Japanese
collaboration that routinely employs TIFR's upgraded Giant Metrewave
Radio Telescope for monitoring a set of IPTA pulsars.  BCJ, YG, YM,
SD, AG and PR acknowledge the support of the Department of Atomic
Energy, Government of India, under Project Identification \# RTI 4002.
BCJ, YG and YM acknowledge support of the Department of Atomic Energy,
Government of India, under project No. 12-R\&D-TFR-5.02-0700 while SD,
AG and PR acknowledge support of the Department of Atomic Energy,
Government of India, under project no. 12-R\&D-TFR-5.02-0200.  KT is
partially supported by JSPS KAKENHI Grant Numbers 20H00180, 21H01130,
and 21H04467, Bilateral Joint Research Projects of JSPS, and the ISM
Cooperative Research Program (2021-ISMCRP-2017). AS is supported by
the NANOGrav NSF Physics Frontiers Center (awards \#1430284 and
2020265).  AKP is supported by CSIR fellowship Grant number
09/0079(15784)/2022-EMR-I.  SH is supported by JSPS KAKENHI Grant
Number 20J20509.  KN is supported by the Birla Institute of Technology
\& Science Institute fellowship.  AmS is supported by CSIR fellowship
Grant number 09/1001(12656)/2021-EMR-I and T-641 (DST-ICPS).  TK is
partially supported by the JSPS Overseas Challenge Program for Young
Researchers.  We acknowledge the National Supercomputing Mission (NSM)
for providing computing resources of ‘PARAM Ganga’ at the Indian
Institute of Technology Roorkee as well as `PARAM Seva' at IIT
Hyderabad, which is implemented by C-DAC and supported by the Ministry
of Electronics and Information Technology (MeitY) and Department of
Science and Technology (DST), Government of India. DD acknowledges the 
support from the Department of Atomic Energy, Government of India 
through Apex Project - Advance Research and Education in Mathematical 
Sciences at IMSc. \fi

The work presented here is a culmination of many years of data
analysis as well as software and instrument development. In particular,
we thank Drs. N.~D'Amico, P.~C.~C.~Freire, R.~van Haasteren, 
C.~Jordan, K.~Lazaridis, P.~Lazarus, L.~Lentati, O.~L\"{o}hmer and 
R.~Smits for their past contributions. We also
thank Dr. N. Wex for supporting the calculations of the
galactic acceleration as well as the related discussions.
\ifnum\wm=5 We would like to thank Prof. Drs. Alexey Starobinskiy, Sergei Blinnikov and Alexander Dolgov for discussions on the early Universe physics. \fi
\ifnum\wm=5 HM acknowledges the support of the UK Space Agency, Grant No. ST/V002813/1 and ST/X002071/1. Some of the computations described in this paper were performed using the University of Birmingham's BlueBEAR HPC service, which provides a High Performance Computing service to the University's research community. See~\url{http://www.birmingham.ac.uk/bear} for more details. \fi
The EPTA is also grateful
to staff at its observatories and telescopes who have made the
continued observations possible.

We also thank the referee for their insightful and timely comments which has improved the quality of the manuscript.
\linebreak\linebreak\textit{Author contributions.}
The EPTA is a multi-decade effort and all authors have
contributed through conceptualisation, funding acquisition,
data-curation, methodology, software and hardware
 developments as well as (aspects of) the continued running of
the observational campaigns, which includes writing and
proofreading observing proposals, evaluating observations
and observing systems, mentoring students, developing
science cases. All authors also helped in (aspects of)
verification of the data, analysis and results as well as
in finalising the paper draft. Specific contributions from individual 
EPTA members are listed in the CRediT\footnote{\url{https://credit.niso.org/}} format below.

InPTA members contributed in uGMRT observations and data reduction to
create the InPTA data set which is employed while assembling the
\texttt{DR2full+} and \texttt{DR2new+} data sets. 

\ifnum\wm=1

JJan, KLi, GMS equally share the correspondence of the paper.

\linebreak\linebreak\textit{CRediT statement:}\newline
Conceptualisation: APa, APo, AV, BWS, CGB, CT, GHJ, GMS, GT, IC, JA, JJan, JPWV, JW, JWM, KJL, KLi, MK.\\
Methodology: APa, AV, DJC, GMS, IC, JA, JJan, JPWV, JWM, KJL, KLi, LG, MK.\\
Software: AC, AJ, APa, CGB, DJC, GMS, IC, JA, JJan, JJaw, JPWV, KJL, KLi, LG, MJK, RK.\\
Validation: AC, APa, CGB, CT, GMS, GT, IC, JA, JJan, JPWV, JWM, KLi, LG.\\
Formal Analysis: APa, CGB, DJC, DP, EvdW, GHJ, GMS, JA, JJan, JPWV, JWM, KLi.\\
Investigation: APa, APo, BWS, CGB, DJC, DP, GMS, GT, IC, JA, JJan, JPWV, JWM, KLi, LG, MBM, MBu, MJK, RK.\\
Resources: APa, APe, APo, BWS, GHJ, GMS, GT, HH, IC, JA, JJan, JPWV, JWM, KJL, KLi, LG, MJK, MK, RK.\\
Data Curation: AC, AJ, APa, BWS, CGB, DJC, DP, EG, EvdW, GHJ, GMS, GT, HH, IC, JA, JJan, JPWV, JWM, KLi, LG, MBM, MBu, MJK, MK, NKP, RK, SChe, YJG.\\
Writing – Original Draft: APa, GMS, JA, JJan, KLi, LG.\\
Writing – Review \& Editing: AC, AF, APa, APo, DJC, EB, EFK, GHJ, GMS, GT, JA, JJan, JPWV, JWM, KLi, MK, SChe, VVK.\\
Visualisation: APa, GMS, JA, JJan, KLi.\\
Supervision: APo, ASe, AV, BWS, CGB, DJC, EFK, GHJ, GMS, GT, IC, JA, JPWV, KJL, KLi, LG, MJK, MK, VVK.\\
Project Administration: APo, ASe, AV, BWS, CGB, CT, GHJ, GMS, GT, IC, JJan, JPWV, JWM, KLi, LG, MK.\\
Funding Acquisition: APe, APo, ASe, BWS, GHJ, GT, IC, JA, JJan, LG, MJK, MK.\\

\fi

\ifnum\wm=2
InPTA members contributed to the discussions that probed the impact of 
including InPTA data on single pulsar noise analysis. Furthermore, they 
provided quantitative comparisons of various noise models, wrote a brief 
description of the underlying \texttt{Tensiometer} package, and helped 
with the related interpretations.

APa, AC, MJK equally share the correspondence of the paper. 

\linebreak\linebreak\textit{CRediT statement:}\newline
Conceptualisation: AC, APa, APo, AV, BWS, CT, GMS, GT, JPWV, JWM, KJL, KLi, MJK, MK.\\
Methodology: AC, APa, AV, DJC, GMS, IC, JWM, KJL, KLi, LG, MJK, MK, SB, SChe, VVK.\\
Software: AC, AJ, APa, APe, GD, GMS, KJL, KLi, MJK, RK, SChe, VVK.\\
Validation: AC, APa, BG, GMS, IC, JPWV, JWM, KLi, LG, MJK.\\
Formal Analysis: AC, APa, BG, EvdW, GHJ, GMS, JWM, KLi, MJK.\\
Investigation: AC, APa, APo, BWS, CGB, DJC, DP, GMS, IC, JPWV, JWM, KLi, LG, MBM, MBu, MJK, RK, VVK.\\
Resources: AC, APa, APe, APo, BWS, GHJ, GMS, GT, IC, JPWV, JWM, KJL, KLi, LG, MJK, MK, RK.\\
Data Curation: AC, AJ, APa, BWS, CGB, DJC, DP, EvdW, GHJ, GMS, JA, JWM, KLi, MBM, MJK, MK, NKP, RK, SChe.\\
Writing – Original Draft: AC, APa, GMS, MJK.\\
Writing – Review \& Editing: AC, AF, APa, APo, BG, EB, EFK, GMS, GT, JA, JPWV, JWM, KLi, MJK, MK, SChe, VVK.\\
Visualisation: AC, APa, GMS, KLi, MJK.\\
Supervision: AC, APo, ASe, AV, BWS, CGB, DJC, EFK, GHJ, GT, JPWV, KJL, LG, MJK, MK, VVK.\\
Project Administration: AC, APo, ASe, AV, BWS, CGB, CT, GHJ, GMS, GT, JPWV, JWM, LG, MJK, MK.\\
Funding Acquisition: APe, APo, ASe, BWS, GHJ, GT, IC, LG, MJK, MK.\\
\fi

\ifnum\wm=3

Additionally, InPTA members contributed to GWB search efforts with 
\texttt{DR2full+} and \texttt{DR2new+} data sets and their interpretations. 
Further, they provided quantitative comparisons of GWB posteriors that 
arise from these data sets and multiple pipelines.

For this work specifically, SChen and YJG equally share the 
correspondence of the paper. 

\linebreak\linebreak\textit{CRediT statement:}\newline
Conceptualisation: AC, APa, APe, APo, ASe, AV, BG, CT, GMS, GT, IC, JA, JPWV, JWM, KJL, KLi, MK.\\
Methodology: AC, APa, ASe, AV, DJC, GMS, JWM, KJL, KLi, LS, MK, SChe.\\
Software: AC, AJ, APa, APe, GD, GMS, KJL, KLi, MJK, RK, SChe, VVK.\\
Validation: AC, APa, ASe, AV, BG, GMS, HQL, JPWV, JWM, KLi, LS, SChe, YJG.\\
Formal Analysis: AC, APa, ASe, AV, BG, EvdW, GMS, HQL, JWM, KLi, LS, MF, NKP, PRB, SChe, YJG.\\
Investigation: APa, APo, ASe, AV, BWS, CGB, DJC, DP, GMS, JWM, KLi, LS, MBM, MBu, MF, PRB, RK, SB, SChe, YJG.\\
Resources: AC, APa, APe, APo, ASe, AV, BWS, GHJ, GMS, GT, IC, JPWV, JWM, KJL, KLi, LG, LS, MJK, MK, RK.\\
Data Curation: AC, AJ, APa, BWS, CGB, DJC, DP, EvdW, GMS, JA, JWM, KLi, MBM, MJK, MK, RK, SChe.\\
Writing – Original Draft: AC, APa, BG, DJC, GMS, JA, KLi, SB, SChe, YJG.\\
Writing – Review \& Editing: AC, AF, APa, APo, ASe, AV, BG, DJC, EB, EFK, GMS, GT, JA, JPWV, JWM, KLi, LS, MBo, MK, NKP, PRB, SChe, VVK, YJG.\\
Visualisation: APa, BG, GMS, KLi, MF, PRB, SChe.\\
Supervision: APo, ASe, AV, BWS, CGB, DJC, EFK, GHJ, GMS, GT, JPWV, KJL, MK, SB.\\
Project Administration: APo, ASe, AV, BWS, CGB, CT, GHJ, GMS, GT, JPWV, JWM, LG, MK, SChe.\\
Funding Acquisition: APe, APo, ASe, AV, BWS, GHJ, GT, IC, JA, LG, MJK, MK, SB.\\
\fi

\end{acknowledgements}
%
%
%
\bibliographystyle{aa} 
\bibliography{epta_wm2} 

\end{document}